\documentclass[
    aps,
    prb,
    reprint,
    10pt,
    amsmath,
    amssymb,
    superscriptaddress,
    floatfix,
    longbibliography
]{revtex4-2}

\usepackage{amsfonts}
\usepackage{amsthm}
\usepackage{mathrsfs}
\usepackage{latexsym}
\usepackage{physics}
\usepackage{graphicx}
\usepackage{xcolor}
\usepackage{bm}
\usepackage{paralist}
\usepackage{enumitem}

\usepackage[
    colorlinks=true,
    linkcolor=blue,
    citecolor=blue,
    urlcolor=blue
]{hyperref}

\newcommand{\br}{\vb{r}}
\newcommand{\bq}{\vb{q}}

\newcommand{\bQ}{\vb{Q}}
\newcommand{\bS}{\vb{S}}
\newcommand{\bD}{\vb{D}}

\begin{document}
\title{Local spin constraints and symmetry in single-$\mathbf Q$ spiral magnets}
\author{Lasin Thaivalappil}
\affiliation{Department of Physics, National Sun Yat-sen University, Kaohsiung 804, Taiwan}

\author{Shin-Ming Huang}
\email{Contact author: shinming@mail.nsysu.edu.tw}
\affiliation{Department of Physics, National Sun Yat-sen University, Kaohsiung 804, Taiwan}
\affiliation{Center for Theoretical and Computational Physics, National Sun Yat-sen University, Kaohsiung 80424, Taiwan}
\affiliation{Physics Division, National Center for Theoretical Sciences, Taipei, 10617, Taiwan}

\date{\today}

\begin{abstract}
The Luttinger--Tisza (LT) method provides a lower bound on the energy of classical spin models but relaxes the local spin-length constraints. For a generic incommensurate single-$Q$ spiral, fixed spin length requires circular polarization on every magnetic sublattice, a condition not guaranteed by equal sublattice weights in the lowest LT eigenmode. We develop a constrained formulation for general bilinear classical spin Hamiltonians with one or two magnetic sublattices. For one sublattice, minimization reduces to a secular equation for the spiral-plane normal. For two sublattices, exact elimination of the relative phase gives a frame-independent energy on the space of two unit normals. We derive symmetry constraints by distinguishing operations that preserve or reverse the Fourier arm and those that preserve or exchange sublattices. For a model with symmetric anisotropy, numerical optimization over both spiral-plane normals yields a lower energy than either constrained pure exchange-sector state, despite the block diagonalization of the linear Hamiltonian. A separate anisotropic two-sublattice benchmark demonstrates that enforcing local circularity can shift the preferred ordering wave vector even when the lowest LT eigenmode has equal sublattice weights.
\end{abstract}
\maketitle

\section{Introduction}

Magnetic spirals are characterized by two quantities: the propagation vector $\mathbf Q$ and the plane in which the moments rotate. Together they fix the chirality, the polarized-neutron response, and the coupling to charge and topology in materials ranging from B20 helimagnets to Weyl semimetal SmAlSi~\cite{Nagamiya1967,Grigoriev2015,Maleyev2004,bihlmayer2007density,Yao2023,thaiv2026,nakano2026}. Both are selected by the competition between symmetric exchange, Dzyaloshinskii--Moriya (DM) interactions, and crystal symmetry. First-principles methods now deliver detailed exchange tensors~\cite{Liechtenstein1987,Sandratskii1998,Tung2011,Sandratskii2017,Mankovsky_2022,Szilva2023}, so a reliable map from an interaction tensor to the spiral it favors is increasingly needed.

The standard starting point is the Luttinger--Tisza (LT) method~\cite{Luttinger1946,LyonsKaplan1960,Broughton1972,Misra1973,Litvin1974}. It replaces the local constraints $\|\mathbf S(\mathbf r)\|=1$ by a single global normalization, which turns the problem into an eigenvalue problem and gives a rigorous lower bound on the classical energy at every wave vector. The bound is tight only if the lowest eigenmode can be realized by unit-length spins. With several magnetic sublattices this fails when the eigenmode weights the sublattices unequally~\cite{Niggemann_2020,SchmidtRichter2022}. Equal weights do not repair it, because a further requirement exists. A single-$Q$ state with no uniform component, at a generic incommensurate $\mathbf Q$, has constant spin length on every site only if each sublattice amplitude is circularly polarized. An anisotropic interaction can instead select an elliptically polarized mode, whose spin length oscillates along the modulation. 
Such an LT eigenvector is not itself a physical hard-spin configuration. If the lowest eigenspace is degenerate, however, an appropriate linear combination may still satisfy the local constraints and reach the LT bound.

This matters because the discrepancy is not confined to the polarization of the mode. The constraint can change the energy landscape in $\mathbf q$, so it can move the ordering vector itself. Existing remedies include generalized LT constructions with sublattice-dependent Lagrange parameters~\cite{LyonsKaplan1960,SchmidtRichter2022}, real-space minimization, and variational spiral frames~\cite{Zare2013,Iqbal2019,muller2019,Ivanov2021,Halloran2023}. These handle the constraint, but they do not expose how the spiral plane and the relative sublattice phase follow from the structure of the exchange tensor and its symmetry.

Understanding single-$Q$ spirals also provides a foundation for studying more complex magnetic textures. Skyrmion crystals and other multiple-$Q$ states involve coupled modulations at several ordering vectors, as illustrated by the discovery of the skyrmion lattice in MnSi and theoretical studies of frustrated triangular-lattice magnets~\cite{Muhlbauer2009,Okubo2012}. More recent work has established that momentum-dependent anisotropic exchange, constrained by crystal symmetry, can select competing single- and multiple-$Q$ configurations in a broad range of magnetic systems~\cite{YambeHayami2022,YambeHayami2023,HayamiYambe2024}. Assessing this competition requires physically admissible reference states: a lowest-energy Fourier eigenmode need not satisfy the local spin-length constraints, and imposing these constraints can modify both its polarization and its preferred ordering vector. The present formulation addresses this issue by determining optimized circular single-$Q$ states with symmetry-constrained interactions and explicit sublattice structure. Their energies, spiral-plane orientations, and relative sublattice phases provide a controlled starting point for subsequent comparisons with multiple-$Q$ textures and for investigating instabilities of spiral order.

We formulate the circular single-$Q$ problem as an energy minimization over spiral-plane normals, eliminating the relative sublattice phase analytically for two-sublattice systems. We derive the constraints imposed by crystal symmetry and obtain exact solutions in selected cases. Our contributions are as follows.

\begin{enumerate}
\item We recast the local length constraints as two algebraic conditions on a complex amplitude, which makes the circular spiral and its gauge freedom explicit. For one sublattice, the problem reduces to a secular equation for the spiral-plane normal.

\item For two sublattices, we eliminate the relative phase exactly. The energy depends only on two unit normals, and a projector form of the sublattice overlap avoids the coordinate singularities of transverse frames at parallel and antiparallel normals. This also gives a blockwise test for whether an LT eigenvector is physical, and exact nonlinear stationarity equations when it is not.

\item We solve two symmetry settings in closed form. Competing intra-sublattice DM interactions produce a finite angle between the two spiral planes. Under sublattice-exchange symmetry with isotropic symmetric exchange in the exchange-adapted basis, a circular spiral reaches the fixed-$\mathbf Q$ LT bound. For general symmetric anisotropy, however, the constrained minimum need not belong to either pure exchange sector, even when sublattice-exchange symmetry block-diagonalizes the linear Hamiltonian. We demonstrate this through a numerical comparison of the optimal two-normal energy with the constrained pure-sector solutions.

\item Building on existing momentum-space symmetry rules~\cite{YambeHayami2022,YambeHayami2023}, we classify the Hamiltonian constraints for one- and two-sublattice systems by distinguishing operations that preserve or reverse the Fourier arm ($R\bQ=\bQ$ or $R\bQ=-\bQ$) and those that preserve or exchange the two sites, for tetragonal and cubic parent groups. We consider $\mathbf Q=(q,q,0)$, for which the possible nontrivial arm-preserving operations include $M_{1\bar10}$, $M_z$, and $C_{2[110]}$. The same procedure can be applied to other wave-vector directions, such as $(q,0,0)$, using the corresponding symmetry operations.

\item In an anisotropic two-sublattice chain, we demonstrate that enforcing local hard-spin constraint can shift the preferred ordering wave vector even when the lowest LT eigenmode has equal sublattice weights.
\end{enumerate}

Determining the magnetic ground state remains challenging even when the classical spin Hamiltonian is known. Competing interactions in frustrated magnets can favor distinct noncollinear configurations with similar energies~\cite{KaplanMenyuk2007}, while numerical optimization must enforce the length of every local spin. Real-space calculations with periodic boundary conditions also restrict the accessible ordering vectors: an incommensurate modulation must be represented by a commensurate approximant, potentially obscuring small shifts of the preferred $\mathbf Q$. These difficulties motivate analytical reductions alongside numerical approaches, including machine-learning methods for exploring magnetic configurations~\cite{Kwon2022}. Here we consider circular single-$Q$ order at generic incommensurate $\mathbf Q$ in systems with one or two magnetic sublattices. By expressing the local constraints through spiral frames and eliminating the relative sublattice phase analytically, we reduce the constrained optimization to the spiral-plane normals, with crystal symmetry providing further simplifications and exact solutions in selected cases. This formulation also allows us to examine how symmetric anisotropy changes the preferred spiral-plane configuration and why the minimum of the constrained energy need not coincide with the lowest LT wave vector. The optimized configurations and energies provide reference states within the circular single-$Q$ manifold. Establishing their stability against anharmonic, multiple-$Q$, or commensurate structures requires additional comparisons. The formulation also provides a starting point for extending the constrained analysis to coupled modulations at several ordering vectors.

The paper is organized as follows. 
Section~\ref{sec:formulation} defines the interaction matrices and local constraints. Section~\ref{sec:reductions} develops the one- and two-sublattice reductions. Section~\ref{sec:symmetry_solutions} presents the symmetry constraints and exact spiral solutions, and Sec.~\ref{sec:num_2sub} applies the formulation to an anisotropic chain. Section~\ref{sec:conclusion} summarizes the results. Appendix~\ref{app:technical} collects the derivations and implementation details. Appendices~\ref{app:pointgroups_one} and \ref{app:pointgroups_two} contain the complete symmetry classifications for one and two magnetic sublattices, respectively, and Appendix~\ref{app:numerics} presents the numerical validation.

\section{Constrained formulation of circular spirals}
\label{sec:formulation}
\subsection{Fourier-space interaction matrices}
\label{sec:hamiltonian}
Consider a periodic crystal described by the following bilinear spin Hamiltonian
\begin{align}
    \begin{split}
    H &= \sum_{\br,\br'} \sum_{\alpha,\beta=x,y,z}  J_{\alpha \beta}(\br-\br') S_{\alpha}(\br)  S_{\beta} (\br') \\
    &= N \sum_{\bq} \sum_{\alpha,\beta} J_{\alpha \beta}(\bq) S_{\alpha}(-\bq) S_{\beta}(\bq),
    \end{split} \label{eq:H_real_and_fourier}
\end{align}
with the Fourier-transform convention
\begin{equation}
    \bS (\bq) = \frac{1}{N}\sum_{\br} \bS(\br) e^{-i\bq \cdot \br},
    \quad
    J_{\alpha \beta} (\bq) = \sum_{\br} J_{\alpha \beta} (\mathbf{r}) e^{-i\bq \cdot \br}.
\end{equation}
Because $\bS (-\bq) = \bS (\bq)^{*}$ and $H$ is real, $J_{\alpha \beta}(\bq) = J^{*}_{\beta \alpha}(\bq)$; reversing $\bq \rightarrow -\bq$ gives in addition $J_{\alpha \beta}(-\bq) = J^{*}_{\alpha \beta}(\bq)$.

For one magnetic sublattice, Hermiticity separates the Fourier-space interaction into a real symmetric part and an imaginary antisymmetric part. These describe symmetric exchange and the DM interaction, respectively~\cite{moriya1960}. Reality makes the former even and the latter odd under $\bq\mapsto-\bq$. The same statement applies separately to each intra-sublattice block in a multi-site basis.

In matrix form, the Hamiltonian is $H=N\sum_{\vb{q}} \vb{S}(\vb{q})^{\dagger} \mathcal{H}(\vb{q}) \vb{S}(\vb{q})$, with the $3\times 3$ matrix $\mathcal{H}(\vb{q})$ being
\begin{equation}
    \mathcal{H}(\vb{q}) = \mathcal{R}(\vb{q}) + i \mathcal{I}(\vb{q}), \label{RI}
\end{equation}
where $\mathcal{R}(\vb{q})$ and $\mathcal{I}(\vb{q})$ are real symmetric and antisymmetric matrices, respectively. The reality of the real-space spin Hamiltonian requires $\mathcal H(-\vb q)=\mathcal H(\vb q)^*$; hence $\mathcal R(-\vb q)=\mathcal R(\vb q)$ and $\mathcal I(-\vb q)=-\mathcal I(\vb q).$
The antisymmetric part is the cross-product matrix of the DM vector $\bD=(D_x,D_y,D_z)^{\intercal}$,
\begin{align}
    \mathcal{I} &= \mqty(0 & - D_z & D_y \\ D_z & 0 & -D_x \\ -D_y & D_x & 0) = [\vb{D}]_{\cross}, \nonumber \\
   &  ([\bD]_{\cross})_{ij} = \sum_m \epsilon_{imj} D_m . \label{eq:crossmatrix}
\end{align}
The cross-product matrix expresses the cross product of two vectors, $\vb{u}$ and $\vb{v}$, by a matrix product $\vb{u} \cross \vb{v} = [\vb{u}]_{\cross} \vb{v}= [\vb{v}]^{\intercal}_{\cross} \vb{u}$. We will regard a geometric vector as a column matrix and hence identify $\vb{u} \vdot \vb{v} = \vb{u}^{\intercal}\vb{v}$.

\subsection{Circular amplitudes and local spin length}
\label{sec:singleQ}
The classical ground state is the configuration $\expval{\bS(\br)}$ that minimizes $\expval{H}$ subject to $\norm{\bS(\br)}=1$ on every site. 
We consider circular single-$Q$ configurations of unit classical spins at generic incommensurate wave vectors. Throughout, a minimum refers to this constrained manifold unless an LT-bound argument explicitly establishes an unrestricted ground state.

The circular single-$Q$ ansatz is
\begin{equation}
    \expval{\bS(\vb{r})} = \vu{s}_{1} \cos(\vb{Q}\cdot\vb{r}) + \vu{s}_{2}\sin(\vb{Q}\cdot\vb{r}) , 
    \label{eq:ansatz}
\end{equation}
where $\bQ$ is the magnetic ordering vector, and $\vu{s}_{1}$ and $\vu{s}_{2}$ are real orthonormal vectors. The magnetic moment has unit magnitude and rotates within the plane spanned by $\vu{s}_{1}$ and $\vu{s}_{2}$ as the position advances along $\bQ$, with spatial period $2\pi/|\bQ|$. 
Its Fourier transform, 
\begin{equation}
    \expval{\bS(\vb{q})} = \frac{1}{2} \left[ \delta_{\vb{q},\vb{Q}}\left( \vu{s}_{1} - i \vu{s}_{2}\right) + \delta_{\vb{q},-\vb{Q}}\left( \vu{s}_{1} + i \vu{s}_{2}\right)  \right], \label{S_q}
\end{equation} 
shows that $\expval{\bS(\vb{Q})}$ and $\expval{\bS(-\vb{Q})}$ always come together as complex conjugates. 
Expressing the spiral state by the complex representation
\begin{equation}
    \psi = \frac{1}{\sqrt{2}} \left( \vu{s}_{1} - i \vu{s}_{2}\right), \label{eq:psi}
\end{equation}
the energy per site is
\begin{align}
    \begin{split}
        E_{\text{1-sub}}    = \psi^{\dagger}\mathcal{H} \psi. \label{eq:E1Q}
    \end{split}
\end{align} 

The energy depends only on the oriented spiral-plane normal, 
\begin{equation}
    \vu{n} = \vu{s}_{1} \cross \vu{s}_{2}= -i\,\psi\cross \psi^{*} .
    \label{eq:s3_from_psi}
\end{equation}
The energy of the symmetric part is
\begin{equation}
\begin{aligned}
\psi^\dagger\mathcal R\psi
&=\frac12\left(
\vu{s}_{1}^\intercal \mathcal R \vu{s}_{1} +
\vu{s}_{2}^\intercal \mathcal R \vu{s}_{2}
\right) \\
& = \frac{1}{2} \left(
\trace \mathcal R - \vu{n}^\intercal \mathcal R \vu{n}
\right), 
\end{aligned}
\end{equation}
using completeness of the orthonormal frame, $\vu{s}_{1} \vu{s}_{1}^\intercal + \vu{s}_{2} \vu{s}_{2}^\intercal = \mathbb{I} - \vu{n} \vu{n}^\intercal $, and that of the antisymmetric part is 
\begin{equation}
\begin{aligned}
\psi^\dagger i[\mathbf D]_{\cross}\psi
&=\frac12\left[
\vu{s}_{1}\cdot(\mathbf D\cross \vu{s}_{2})
-\vu{s}_{2} \cdot(\mathbf D\cross \vu{s}_{1})
\right] \\
&=-\mathbf D\cdot \vu{n}.
\end{aligned}
\end{equation}
Combining these contributions gives
\begin{equation}
	E_{\text{1-sub}}    = \frac{1}{2} \left(
\trace \mathcal R - \vu{n}^\intercal \mathcal R \vu{n}
\right) -\mathbf D\cdot \vu{n}.
 \label{eq:E1Q_2}
\end{equation} 

The complex representation is convenient for three reasons. (i) The hard-spin constraint is algebraic: Eq.~(\ref{eq:ansatz}) has unit length on every site if and only if
\begin{equation}
    \psi^{\dagger}\psi = 1 \qquad  \mathrm{and}  \qquad \psi^{\intercal}\psi = 0 .
    \label{eq:hard_complex}
\end{equation}
The first equality fixes the sum of the squared lengths of the two real vectors ($\abs{\vu{s}_{1}}^2+\abs{\vu{s}_{2}}^2=2$), while the real and imaginary parts of the second impose equal length ($\abs{\vu{s}_{1}}=\abs{\vu{s}_{2}}$) and orthogonality ($\vu{s}_{1} \cdot \vu{s}_{2} =0$). These conditions are sufficient for any $\bQ$ and necessary for incommensurate $\bQ$; for a commensurate $\bQ$ only finitely many phases $\bQ\vdot\br$ occur and a weaker condition may suffice. (ii) The energy is invariant under the common phase transformation $\psi\mapsto e^{i\gamma}\psi$, corresponding to a shift of the phase origin of the spiral. For a lattice translation $\vb a$ the phase shift is specifically $\gamma=-\bQ\cdot\vb a$; for a generic incommensurate modulation the continuous phase freedom is naturally interpreted as an internal translation of the modulation. (iii) The gauge-invariant content of $\psi$ is precisely the normal $\vu{n}$. The frame--amplitude dictionary is collected in Table~\ref{tab:dictionary}.

\begin{table}[tbp]
\caption{Dictionary between the real orthonormal-frame and the complex
amplitude descriptions of a single-$Q$ spiral.}
\label{tab:dictionary}
\begin{ruledtabular}
\begin{tabular}{ll}
Real frame & Complex amplitude \\
\hline
$\norm{\vu{s}_1}=\norm{\vu{s}_2}=1$, $\vu{s}_1\vdot\vu{s}_2=0$ &
$\psi^{\dagger}\psi=1$, $\psi^{\intercal}\psi=0$ \\
$\vu{n}=\vu{s}_1\cross\vu{s}_2$ & $\vu{n}=-i\,\psi\cross\psi^{*}$ \\
frame rotation by $\gamma$ about $\vu{n}$ & $\psi\mapsto e^{-i\gamma}\psi$ \\
spiral translation by $\vb{a}$ & $\psi\mapsto e^{-i\bQ\vdot\vb{a}}\psi$ \\
$E^{1Q}$ & $\psi^{\dagger}\mathcal H\psi$ \\
\end{tabular}
\end{ruledtabular}
\end{table}

The minimizer of Eq.~(\ref{eq:E1Q}) subject to Eq.~(\ref{eq:hard_complex}) is not in general an eigenvector of $\mathcal H$, as discussed in
Secs.~\ref{sec:exact}--\ref{sec:eigenvector_relation}.

\section{Analytical reduction of the constrained problem}
\label{sec:reductions}
\subsection{One sublattice: a secular problem}
\label{sec:one_sub_theory}
We first consider isotropic symmetric exchange together with a DM interaction, $\mathcal{R} = J  \mathbb{I}$. Let $\vu{D}=\vb{D}/\norm{\vb{D}}$. 
Choose a unit vector $\vu e_1\perp\vb D$ and set $\vu e_2=\vu D\cross\vu e_1$, so that $(\vu e_1,\vu e_2,\vu D)$ is a right-handed orthonormal frame. 
The three eigenvectors (and corresponding eigenvalues) are 
\begin{equation}
\begin{aligned}
\psi_0&=\vu{D} \quad  & (E_{0} = J),\\
\psi_{+} &= \frac{1}{\sqrt{2}} \left( \vu{e}_{1} + i \vu{e}_{2}\right) \quad  &(E_{+} = J + \norm{\vb D}), \\
\psi_{-} &= \frac{1}{\sqrt{2}} \left( \vu{e}_{1} - i \vu{e}_{2}\right) \quad  &(E_{-} = J - \norm{\vb D}),
\end{aligned}
\end{equation}
respectively. 
The lowest mode, $\psi_-$, has precisely the circular form of Eq.~(\ref{eq:psi}) and minimizes the fixed-$\bQ$ energy at $E_{\text{1-sub}}^{\min}=J-\norm{\vb D}$, giving $\vu n= \vu D$. When $\vb D=0$, the spin plane is arbitrary.
With symmetric anisotropy, the lowest LT eigenvectors of $\mathcal H$ need not be circular; in general, $[\mathcal R,\mathcal I]\neq0$.

For a general $\mathcal{H}=\mathcal{R}+i\mathcal{I}$, we diagonalize the symmetric part, 
\begin{equation}
    \mathcal{R} = \sum_{i=1}^3 h_i \vu{z}_i \vu{z}_i^{\intercal}, 
    \qquad h_1 \le h_2 \le h_3 ,
    \label{eq:R_diag}
\end{equation}
and expand the DM vector and the normal in the same basis as 
\begin{equation}
\begin{aligned}
	\bD=\sum_i D_i\vu{z}_i, \qquad
	\vu{n}=\sum_i\alpha_i\vu{z}_i
\end{aligned}
\end{equation}
with $\sum_i\alpha_i^2=1$, which is constrained by a Lagrange multiplier, say $\lambda $. The modified energy becomes
\begin{equation}
    E_{\text{1-sub}}' = \sum_{i=1}^{3} \left[ \frac{1}{2} h_i (1-\alpha_i^2) - D_i \alpha_i\right] + \frac{1}{2} \lambda \left(\sum_{i=1}^{3} \alpha_i^2 -1 \right), \label{eq:E_1_lagrange}
\end{equation}
the factor $1/2$ in front of $\lambda$ being convenient below. 
The stationary condition is
\begin{equation}
	\pdv{E_{\text{1-sub}}'}{\alpha_i}  = (\lambda-h_i)\alpha_i-D_i=0.
	\label{eq:dEda0}
\end{equation}

\begin{figure}[tb]
\centering
\includegraphics[width=0.45\textwidth]{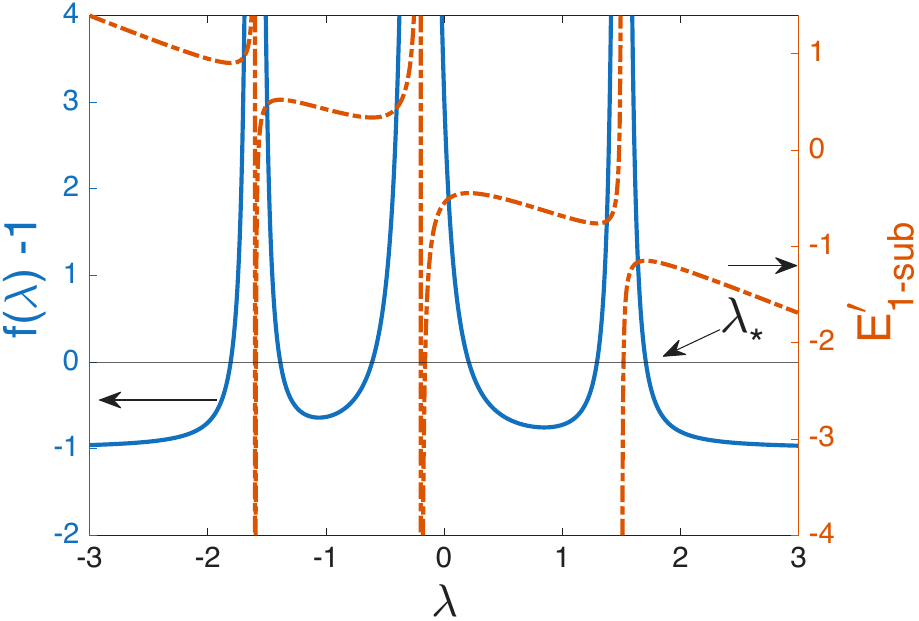}
\caption{Constraint function $f(\lambda)-1$ (blue solid line) and the Lagrangian $E'_{\text{1-sub}}(\lambda)$ (red dash-dotted line) for $(h_1,h_2,h_3)=(-1.6,-0.2,1.5)$ and $(D_1,D_2,D_3)=(0.2,-0.4,0.2)$. The poles occur at $\lambda=h_i$. For the nonzero $D_i$ used here, only the zeros of $f(\lambda)-1$ correspond to constrained stationary states. The global minimum lies on the regular branch at the largest root, $\lambda_*=1.704953$. } \label{fig:lambda}
\end{figure}

When $\vb D\neq0$, we have a simple expression for the spin components 
\begin{equation}
    \alpha_i = \frac{D_i}{\lambda - h_i} \quad (i=1,2,3),  \label{eq:ai}
\end{equation} 
($\lambda \neq h_i$) and the constraint becomes the secular equation
\begin{equation}
    f(\lambda) \equiv\ \sum_{i} \left( \frac{D_i}{\lambda - h_i} \right)^2 = 1. 
     \label{eq:lambda}
\end{equation}
When $D_i\neq0$ for every $i$, $f(\lambda)$ vanishes as $|\lambda|\to\infty$ and diverges at each distinct eigenvalue $h_i$. There is one root on each exterior interval, and additional roots may occur between the poles. 
To select the physical root, rewrite Eq.~(\ref{eq:E_1_lagrange}) by expanding the squares,
 \begin{equation}
    \begin{split}
        E_{\text{1-sub}}' =& \frac{1}{2} \left[ \left( \sum_{i=1}^{3} h_i \right) - \lambda \right] - \frac{1}{2}  \sum_{i=1}^{3} \frac{D_i^2}{\lambda - h_i} \\ &+ \frac{1}{2}  \sum_{i=1}^{3} \left(\lambda - h_i \right) \left(\alpha_i - \frac{D_i}{\lambda - h_i} \right)^2 ,
    \end{split}
\end{equation}
The last term vanishes by Eq.~(\ref{eq:ai}), and the remainder obeys $\pdv*{E_{\text{1-sub}}'}{\lambda} =\tfrac12\left[f(\lambda)-1\right]$. The energy therefore decreases with $\lambda$ wherever $f<1$, which is the case outside the neighbourhoods of the poles, so the global minimum is obtained at the largest root $\lambda_*\ > h_ 3$ of Eq.~(\ref{eq:lambda}). If a regular root exists above $h_3$, it is regular and unique there. 
Figure~\ref{fig:lambda} illustrates the regular construction.
For a regular root, the result can be written in matrix form:
\begin{align}
 \vu{n} &=(\lambda_* \mathbb I-\mathcal{R})^{-1}\vb D,\nonumber\\
 E_{\text{1-sub}} &= \frac{1}{2} \left[ \trace \mathcal{R}-\lambda_*
 -\vb D ^\intercal (\lambda_* \mathbb I-\mathcal{R})^{-1}\vb D \right].
\end{align}
For a regular solution, $\alpha_i=D_i/(\lambda_*-h_i)$ with $\lambda_*>h_3$, so each nonzero component has the sign of $D_i$. The denominators favor the maximal-eigenvalue direction, while the numerators encode the DM orientation. 
In other words, the symmetric interaction favors a normal in its maximal eigenspace, while the DM term favors alignment with $\vb D$; their competition determines the spiral plane.

When $\vb D$ has no component in $\vu{z}_3$, Eq.~(\ref{eq:dEda0}) suggests that the singular minimum at $\lambda_*=h_3$. It leads to $\alpha_i=D_i/(h_3-h_i)$ for $i$ which $h_i$ is smaller than $h_3$. 
To satisfy the secular equation in Eq. (\ref{eq:lambda}), the normal becomes
\begin{equation}
\begin{aligned}
 \vu n&=\sum_{h_i<h_3}\frac{D_i}{h_3-h_i}\vu z_i+\vb v,
 \qquad \vb v\in \text{eigenspace of } h_3,\\
 \abs{ \vb v}^2&=1-\sum_{h_i<h_3}\frac{D_i^2}{(h_3-h_i)^2}.
\end{aligned}
 \label{eq:singular_complete}
\end{equation}
It exists when the right-hand side of $\abs{\vb v}^2$ is nonnegative. (When the lower-eigenspace sum exceeds one, the root returns to a regular one lying above $h_3$.)
Its energy is $\tfrac12[\operatorname{Tr}\mathcal R-h_3-\sum_{h_i<h_3}D_i^2/(h_3-h_i)]$.
This includes the $\vb {D=0}$ case: 
If $h_2 \neq h_3$, $\vec{\alpha}  =(\alpha_1,\alpha_2,\alpha_3)= (0,0,\pm1)$. If $h_2 = h_3 > h_1$, then $\alpha_1=0$ and $\alpha_2,\alpha_3$ are arbitrary subject to normalization. If all $h$'s are equal, then $\vec{\alpha} $ is arbitrary subject to normalization.

Table~\ref{tab:principal_axis} gives the explicit solution when the DM vector along one principal axis, $\vb D=D \vu{z}_i$ ($i=1,2,3$). 
At a degeneracy of the maximal eigenvalue, the transverse component can be any vector of the prescribed length in the maximal eigenspace orthogonal to the DM axis. 
The discussion of these cases is especially important when symmetries are considered which give $\mathcal{D}_g \vu{z}_i =  \vu{z}_i$ while $\mathcal{D}_g \vu{z}_j =  -\vu{z}_j ~(j\neq i)$ as discussed in Sec.~\ref{sec:one_symmetry_solution}. 
\begin{table*}[t]
\caption{One-sublattice minima for $\bD=D\vu z_i$, $D\ne0$, and $h_1\leq h_2\leq h_3$. Write $\Delta_i=h_3-h_i$. The displayed tilted vectors assume a nondegenerate maximal eigenvalue; the text specifies the degenerate extension.}
\label{tab:principal_axis}
\begin{ruledtabular}
\begin{tabular}{lllll}
Axis & Regime & $\lambda_*$ & $ \vu{n} $ & $E_{\text{1-sub}}^{\min}$\\
\hline
$i=1,2$ & $0<|D|<\Delta_i$ & $h_3$ & $(D/\Delta_i)\vu z_i\pm\sqrt{1-D^2/\Delta_i^2}\vu z_3$ & $(h_1+h_2)/2-D^2/(2\Delta_i)$\\
$i=1,2$ & $|D|\geq\Delta_i$ & $h_i+|D|$ & $\mathrm{sgn}(D)\vu z_i$ & $\tfrac12\sum_{j\ne i}h_j-|D|$\\
$i=3$ & all $D\ne0$ & $h_3+|D|$ & $\mathrm{sgn}(D)\vu z_3$ & $(h_1+h_2)/2-|D|$\\
\end{tabular}
\end{ruledtabular}
\end{table*}

\subsection{Two sublattices: constrained stationarity}
\label{sec:two_sub_theory}
\label{sec:exact}

The one-sublattice geometry now provides the local building block for two magnetic sites per primitive cell. We first impose both local constraints, then eliminate the relative phase. 
With $\mu=1,2$ labelling the two basis sites, define
\begin{equation}
 \Psi=\begin{pmatrix}\psi^{(1)}\\ \psi^{(2)}\end{pmatrix},\qquad
 \psi^{(\mu)}=\frac{1}{\sqrt2}
 \left(\vu{s}^{(\mu)}_1-i\vu{s}^{(\mu)}_2\right).
 \label{eq:Psi_def}
\end{equation}
so that the real-space spins are
\begin{equation}
 \vb S^{(\mu)}(\vb R)=\sqrt2\,\Re\!\left[
 \psi^{(\mu)}e^{i\vb Q\cdot(\vb R+\vb r_\mu)}\right].
 \label{eq:realspace_reconstruction}
\end{equation}
where $\vb R$ is a Bravais-lattice vector and $\vb r_\mu$ is a basis position. Sublattice-dependent orthonormal frames are established variational coordinates for spiral states~\cite{Halloran2023}. Here normalization and circularity provide an equivalent algebraic description of the same local frame geometry. 

The interaction matrix and its diagonal blocks are
\begin{equation}
    \mathcal{H}_{\text{2-sub}} = \mqty(\mathcal{H}_{11} & \mathcal{H}_{12} \\ \mathcal{H}_{12}^\dagger & \mathcal{H}_{22}),
    \label{eq:H2sub}
\end{equation}
\begin{align}
        \mathcal{H}_{\mu \mu} =& \sum_{i=1}^3 h_i^{(\mu)} \vu{z}_i^{(\mu)} \vu{z}_i^{{(\mu)} \intercal} + i [\vb{D}^{(\mu)}]_{\cross},
        \label{eq:Hmumu}
\end{align}
The energy $E_{\text{2-sub}}=\Psi^\dagger\mathcal H_{\text{2-sub}}\Psi$ is expressed per primitive cell, whereas $E_{\text{1-sub}}$ is per site. The inter-sublattice block $\mathcal H_{12}$ is arbitrary complex and $\mathcal H_{21}=\mathcal H_{12}^\dagger$.

Introducing a real multiplier $\varepsilon_\mu$ and a complex multiplier $\kappa_\mu$ for the two constraints on sublattice $\mu$, the constrained energy is
\begin{align}
 E_{\text{2-sub}}'={}&\Psi^\dagger\mathcal H_{\text{2-sub}}\Psi
 -\sum_\mu\varepsilon_\mu(\psi^{(\mu)\dagger}\psi^{(\mu)}-1)\nonumber\\
 &-\Re\sum_\mu\kappa_\mu^*\psi^{(\mu) \intercal }\psi^{(\mu)}.
 \label{eq:L_complex}
\end{align}
The stationary solution satisfies the nonlinear eigenvalue equations
\begin{equation}
 \quad
 \sum_{\nu=1}^2\mathcal H_{\mu\nu}\psi^{(\nu)}
 =\varepsilon_\mu\psi^{(\mu)}+\kappa_\mu\psi^{(\mu)*},
 \quad \mu=1,2.\quad
 \label{eq:nonlinear_eigen}
\end{equation}
The problem is equivalent to the zero-eigenvalue equation: 
\begin{equation}
 \begin{pmatrix}
 \mathcal H-\mathcal E&-K\\
 -K^*&\mathcal H^*-\mathcal E
 \end{pmatrix}\Phi=0,
 \label{eq:doubled}
\end{equation}
where $\Phi=(\Psi,\Psi^*)$, $\mathcal E = \mathrm{diag}(\varepsilon_1\mathbb I_3, \varepsilon_2\mathbb I_3)$, and $K=\mathrm{diag}(\kappa_1\mathbb I_3, \kappa_2\mathbb I_3)$.
The physical eigenvector of zero-eigenvalue is achieved by tuning the six real constraints (two real $\varepsilon_\mu$ and two complex $\kappa_\mu$).

The amplitudes $\psi^{(\mu)}$ and $\psi^{(\mu)*}$ describe opposite circular polarizations. 
The term proportional to $\kappa_\mu$ describes the component along the opposite circular polarization.  
Multiplication by $\psi^{(\mu)\dagger}$ and $\psi^{(\mu) \intercal }$ yields, respectively,
\begin{align}
 \varepsilon_\mu&=\psi^{(\mu)\dagger}
 \sum_\nu\mathcal H_{\mu\nu}\psi^{(\nu)}, \label{eq:eps_extract}\\
 \kappa_\mu&=\psi^{(\mu) \intercal }
 \sum_\nu\mathcal H_{\mu\nu}\psi^{(\nu)}. \label{eq:kappa_extract}
\end{align}
Summing Eq.~(\ref{eq:eps_extract}) over $\mu$ shows that the stationary energy equals the sum of the real multipliers: $E=\Psi^\dagger\mathcal H_{\text{2-sub}}\Psi= \varepsilon_1+\varepsilon_2$. 

Equations (\ref{eq:nonlinear_eigen}) and (\ref{eq:hard_complex}) constitute the exact stationarity conditions for the two-sublattice circular single-$Q$ hard-spin problem. Together, they form a finite system of polynomial equations in the real and imaginary parts of the six amplitude components and the associated multipliers. In contrast to the standard linear eigenvalue problem, this formulation retains the local unit-length constraint explicitly.

The doubled algebraic representation is given in Appendix~\ref{app:technical}. For numerical work, feasible orthonormal-frame updates preserve both constraints exactly, and the reconstructed amplitudes can be tested against Eq.~(\ref{eq:nonlinear_eigen}).

\subsection{Relation to eigenvectors and the Luttinger-Tisza bound}
\label{sec:eigenvector_relation}

Let $v_n=(v_n^{(1)},v_n^{(2)})^{\intercal}$ be a normalized eigenvector of $\mathcal H(\bQ)$ with eigenvalue $E_n$. The rescaled eigenvector $\Psi=\sqrt2\,v_n$ represents a hard-spin single-$Q$ state if and only if (for equal unit moments)

\begin{equation}
 v_n^{(\mu)\dagger}v_n^{(\mu)}=1/2 \qquad  \mathrm{and}  \qquad 
 v_n^{(\mu)\intercal}v_n^{(\mu)}=0
 \qquad (\mu=1,2).
 \label{eq:eigenvector_test}
\end{equation}
The first condition fixes equal spin lengths on the two sublattices; the second states that each sublattice block is circularly polarized. For a generic incommensurate $\bQ$, these two blockwise conditions are therefore necessary and sufficient. 
When a lowest eigenvector satisfies Eq.~(\ref{eq:eigenvector_test}), it defines a physical circular spiral whose energy equals the LT lower bound at the chosen $\bQ$. If $\bQ$ also globally minimizes $\lambda_{\min}[\mathcal H(\bq)]$, this configuration reaches the global LT lower bound and is therefore an exact classical ground state.

If the lowest eigenvalue is nondegenerate and its eigenvector fails Eq.~(\ref{eq:eigenvector_test}), the eigenvector itself is not a physical hard-spin configuration. When the lowest eigenspace is degenerate, however, a linear combination may restore the constraints. 
If no vector in the lowest eigenspace satisfies these constraints, the exact spiral necessarily contains higher-eigenvalue components. Equation~(\ref{eq:nonlinear_eigen}) quantifies the admixture: the anomalous term $\kappa_\mu\psi^{(\mu)*}$ and the inequality of the two multipliers are exactly what prevents $\Psi$ from being an ordinary eigenvector.

\subsection{Exact elimination of the relative phase}
\label{sec:two_sub}\label{sec:general_ABC}

Unlike the intra-sublattice energy, which depends only
on the spiral-plane normals $\vu n_\mu$, the
inter-sublattice interaction also depends on the relative
orientation of the in-plane spin vectors.

For each sublattice, choose an orthonormal reference
frame $\{\vu s_1^{\prime(\mu)},\vu s_2^{\prime(\mu)}\}$
perpendicular to the corresponding normal $\vu n_\mu$.
The physical in-plane frames are obtained by rotations
through angles $\phi^{(\mu)}$, parameterized as

\begin{equation}
	\mqty(\vu{s}_{1}^{(\mu)} \\ \vu{s}_{2}^{(\mu)}) = \mqty(\cos\phi^{(\mu)} & -\sin\phi^{(\mu)} \\ \sin\phi^{(\mu)} & \cos\phi^{(\mu)}) 
	\mqty(\vu{s}_{1}^{\prime (\mu)} \\ \vu{s}_{2}^{\prime (\mu)}),
    \label{eq:frame_rotation}
\end{equation}
The transformation is equivalent to $\psi^{(\mu)}=e^{-i\phi^{(\mu)}}\psi^{\prime(\mu)}$ with $\psi^{\prime(\mu)}=(\vu{s}_{1}^{\prime(\mu)} -i\vu{s}_{2}^{\prime(\mu)})/\sqrt{2}$. The inter-sublattice energy therefore becomes 
\begin{equation}
    E^{\text{inter}}_{\text{2-sub}}
    = 2\,\Re\!\left(e^{-i\Delta\phi}\,\mathcal Z\right) 
    \label{eq:Z_def}
\end{equation}
where 
\begin{equation}
	\mathcal Z \equiv \psi^{\prime(1)\dagger}\mathcal H_{12} \psi^{\prime(2)} , \label{Z}
\end{equation}
and \(\Delta\phi=\phi^{(2)}-\phi^{(1)}\). 
Writing $\mathcal Z=|\mathcal Z|e^{i\vartheta}$ gives $E^{\mathrm{inter}}=2|\mathcal Z|\cos(\vartheta-\Delta\phi)$. Minimization therefore gives
\begin{equation}
 E^{\mathrm{inter}}_{\min}=-2|\mathcal Z|.
 \label{eq:Emin_Z}
\end{equation}
For $\mathcal Z\neq0$, a minimizing phase is
\begin{equation}
 \Delta\phi=\arg\mathcal Z+\pi\pmod{2\pi}.
 \label{eq:phase_solution_general}
\end{equation}
When $\mathcal Z=0$, the relative phase is irrelevant. This elimination is the scalar phase-optimization step~\cite{Higham1986}.

The overlap modulus can be evaluated without constructing intersecting spin planes. For any normalized circular amplitude, define
\begin{equation}
 \Pi(\vu n)=\psi\psi^\dagger
 =\frac12\left(\mathbb I-\vu n\vu n^T-i[\vu n]_{\cross}\right).
 \label{eq:main_projector}
\end{equation}
Then, for arbitrary $\mathcal H_{12}$,
\begin{equation}
 |\mathcal Z|^2=\operatorname{Tr}
 [\Pi(\vu n_1) \mathcal H_{12} \Pi(\vu n_2) \mathcal H_{12}^\dagger].
 \label{eq:main_overlap_trace}
\end{equation}
This identity follows by multiplying the two overlap factors and using the trace of a rank-one matrix. It is valid for parallel and antiparallel normals as well as for tilted planes.

After phase minimization, the physical energy is
\begin{equation}
\begin{aligned}
E_{\text{2-sub}}={}&\sum_{\mu=1}^2\left[
\frac12\left(\trace\mathcal R^{(\mu)}
-\vu n_\mu^\intercal\mathcal R^{(\mu)}\vu n_\mu\right)
-\vb D^{(\mu)}\cdot\vu n_\mu\right]\\
&-2|\mathcal Z(\vu n_1,\vu n_2)|.
\end{aligned}
\label{eq:E2sub_final}
\end{equation}
The optimal energy turns out to depend only on the two spiral-plane normals. The remaining search is therefore on $S^2\times S^2$. Appendix~\ref{app:frames} supplies explicit frame coordinates and component formulas; the coupled normal equations are also shown. 

\section{Symmetry and selection of spiral configurations}
\label{sec:symmetry_solutions}
\subsection{Arm and sublattice actions}
\label{sec:one_sub_little}
Let $G_{\bQ}$ denote the (unitary) little group of the parent Hamiltonian at $\bQ$,
whose symmetry operations satisfy $R_g\bQ=\bQ+\vb G$, with $\vb G$ a
reciprocal-lattice vector. For a generic $\bQ$ not at the border of the Brillouin zone, the relevant operations satisfy
$R_g\bQ=\bQ$ exactly. The magnetic order might break the symmetry $g$; it need not be preserved by the
ordered spiral.

For $g=\{R_g|\boldsymbol\tau_g\}$ preserving the single-site magnetic
lattice, its representation at $\bQ$ is
$\mathcal U_g =e^{-i \bQ \cdot \boldsymbol\tau_g}\mathcal{D}_g$, where
$\mathcal{D}_g=\det(R_g)R_g$ is the axial spin matrix, which is orthogonal [$\det(R_g)$ appears because spin is an axial vector]; for inversion, mirrors, or twofold rotations considered below, $\mathcal{D}_g^\intercal=\mathcal{D}_g$ and $\mathcal{D}_g^2=\mathbb I$. The scalar Bloch phase
cancels from $\mathcal{U}_g\mathcal{H} \mathcal{U}_g^\dagger=\mathcal{H}$. Thus, with
$\mathcal H=\mathcal R+i[\bD]_{\cross}$,
\begin{equation}
 \mathcal{D}_g\mathcal R \mathcal{D}_g^\intercal =\mathcal R,\qquad \mathcal{D}_g\bD=\bD.
 \label{eq:one_little_constraints}
\end{equation}
These are the one-sublattice form of momentum-space exchange-symmetry rules discussed in Refs.~\cite{YambeHayami2022,YambeHayami2023}. They follow from
$\mathcal{D}_g[\bD]_{\cross} \mathcal{D}_g^\intercal=[\mathcal{D}_g\bD]_{\cross}$, since $\det \mathcal{D}_g=+1$.
All quantities in this discussion are evaluated at the same $\bQ$.

If the parent space group contains an operation $g_-$ that reverses $\bQ$, the group preserving the pair of arms is $G_{\{\pm\mathbf Q\}}=G_{\mathbf Q}\cup g_-G_{\mathbf Q}$~\cite{PerezMato2012}. Operations in $G_{\mathbf Q}$ preserve the arm, $R_g\bQ=\bQ$, whereas those in the second coset reverse it, $R_{g'}\bQ=-\bQ$.
An arm-reversing spatial operation acts as $g': \psi\mapsto e^{i\beta_{g'}}\mathcal{D}_{g'}\psi^*$; conjugation here comes from the reality relation between the two Fourier arms. 
The Hamiltonian is thus constrained by $g'$ as
\begin{equation}
\begin{aligned}
\mathcal D_{g'} \mathcal H(\mathbf Q)^*  \mathcal D_{g'}^\intercal &=\mathcal H(\mathbf Q), \\
 \mathcal{D}_{g'}\mathcal R \mathcal{D}_{g'}^\intercal &=\mathcal R,& \mathcal{D}_{g'}\bD&=-\bD.
 \label{eq:one_anti_constraints}
\end{aligned}
\end{equation}
Note that $H(-\bQ)=H(\bQ)^*$ as we explained in Sec. \ref{sec:hamiltonian}.

For the tetragonal and cubic settings considered here, the possible arm-preserving operations at a generic $\bQ=(q,q,0)$ are $E$, $M_{1\bar10}$, $M_z$, and $C_{2[110]}$, while the possible arm-reversing operations are $I_{\text{nv}}$, $C_{2[1\bar10]}$, $C_{2z}$, and $M_{110}$. The operations actually present depend on the parent point group; the corresponding subsets are listed in Appendix~\ref{app:pointgroups_one}.

The right-handed coordinate basis used below is
\begin{equation}
 \hat{\vb e}_{\parallel}=\frac{(1,1,0)^\intercal}{\sqrt2},\quad
 \hat{\vb e}_{\perp}=\frac{(-1,1,0)^\intercal}{\sqrt2},\quad
 \hat{\vb e}_z=(0,0,1)^\intercal.
 \label{eq:diagonal_exchange_basis}
\end{equation}
For a two-site basis, the site permutation supplies a second, independent action. Table~\ref{tab:two_arm_block_constraints} summarizes the four combinations. An exchanging arm-reversing operation constrains $H_{12}$ through its transpose, whereas an exchanging arm-preserving operation uses its Hermitian conjugate. The distinction permits opposite intra-sublattice DM fields under exchanging inversion. The derivation is given in Appendix~\ref{sec:two_sub_arm_reversing}.

The site actions must satisfy the full parent-group relations. With exactly two magnetic sites per primitive cell, $C_{2z}$ preserves both sites in every tetragonal and cubic group listed in Appendix~\ref{app:pointgroups_two}. In the tetragonal case this follows because it is the square of a fourfold operation, while any two-site permutation squares to the identity. The complete classification yields one, two, or four fixed-$\bQ$ cases depending on the parent group, rather than independent choices for every symmetry operation.

\begin{table*}[t]
\caption{Hamiltonian constraints at fixed $\bQ$ for the four combinations of arm and sublattice actions. $\mathcal D$ is real orthogonal for a symmetry operation. For exchanging operations, the compact diagonal-block relation shown assumes $\mathcal D^2=\mathbb I$, as for the inversion, mirrors, and twofold rotations considered here. Common translation phases have been canceled in the position-based Fourier gauge.}
\label{tab:two_arm_block_constraints}
\begin{ruledtabular}
\begin{tabular}{llll}
Fourier arm & Sublattices & Diagonal blocks & Inter-sublattice block\\
\hline
Preserved & Preserved & $\mathcal H_{\mu\mu}=\mathcal D \mathcal H_{\mu\mu}\mathcal D^\intercal$ & $H_{12} =\mathcal D \mathcal H_{12} \mathcal D^\intercal$\\
Preserved & Exchanged & $\mathcal H_{22}=\mathcal D \mathcal H_{11}\mathcal D^\intercal$ & $H_{12}=\mathcal D H_{12}^\dagger\mathcal D^\intercal$\\
Reversed & Preserved & $\mathcal H_{\mu\mu}=\mathcal D \mathcal H_{\mu\mu}^*\mathcal D^\intercal$ & $H_{12}=\mathcal D H_{12}^*\mathcal D^\intercal$\\
Reversed & Exchanged & $\mathcal H_{22}=\mathcal D \mathcal H_{11}^*\mathcal D^\intercal$ & $H_{12}=\mathcal D H_{12}^\intercal\mathcal D^\intercal$\\
\end{tabular}
\end{ruledtabular}
\end{table*}
\subsection{One-sublattice plane selection}
\label{sec:one_symmetry_solution}
The Hamiltonians for the three symmetries are similar. For example, in the presence of $C_{2[110]}$, the Hamiltonian is written as 
\begin{equation}
\mathcal H_{C_{2[110]}} =\mqty(\mathcal R_{\parallel \parallel} & 0 & 0 \\ 0 & \mathcal R_{\perp \perp} &  \mathcal R_{\perp z} - iD \\ 0 & \mathcal R_{\perp z} + iD & \mathcal R_{zz}).
\end{equation}
for $\mathcal D_g = \mathrm{diag}(+1,-1,-1)$.
The other two Hamiltonians are obtained by permuting the basis: $\hat{\vb e}_\parallel \to \hat{\vb e}_\perp \to \hat{\vb e}_z \to \hat{\vb e}_\parallel$. 
After diagonalizing the symmetric transverse block, each of these Hamiltonians can be written as
\begin{equation}
 \mathcal R=\mathrm{diag}(h_a,h_b,h_c),\qquad
 \bD=D \vu a,\qquad h_b\leq h_c.
 \label{eq:one_involution_principal}
\end{equation}
Here $\vu a$ is the axis which is invariant under $\mathcal D_g$, while the other two $\vu b, \vu c$ change sign; $h_a$ is not assumed to be the
largest eigenvalue. We define $ \Delta=h_c-h_a$.

For $D \neq 0$, referring to the results in Sec.~\ref{sec:one_sub_theory}, the minimal energy is
\begin{equation}
 E_{\text{1-sub}}=
 \begin{cases}
 (h_a+h_b)/2-D^2/(2\Delta),&\Delta>\abs{D} >0,\\
 (h_b+h_c)/2-\abs{D},&\text{otherwise}.
 \end{cases}
 \label{eq:one_symmetry_energies}
\end{equation}
The corresponding $\vu n$ is
\begin{equation}
 \vu n =
 \begin{cases}
 \frac{D}{\Delta}\hat{\vb a} \pm\sqrt{1-\frac{D^2}{\Delta^2}}\,\vu{c} ,& \Delta>\abs{D} >0,\\
 \mathrm{sgn}(D) \vu a ,&\text{otherwise}.
 \end{cases}
 \label{eq:one_tilted_domains}
\end{equation}
If $h_b=h_c$, $\vu c$ can be replaced by any unit transverse direction in the $b$-$c$ plane.
When $\abs{D} > \Delta$, the DM interaction will pin the normal $\vu n$ aligned with the symmetry axis $\vu a$. 
On the other hand, when $\Delta>\abs{D} >0$, symmetric anisotropy favors a transverse component of the normal that competes with the DM term. 
Since $\vu a$ and $\vu c$ take opposite eigenvalues of $\mathcal D_g$, the tilted solution breaks the corresponding symmetry, even up to a modulation phase.

For $D=0$, the normal lies in the maximal eigenspace of $\mathcal R$:
$\vu n = \pm \vu a$ if $h_a > h_c$; $\vu n = \pm \vu c$ if $h_c >h_a$; and for $h_a = h_c$ any unit vector in the
combined maximal eigenspace is allowed.

\subsection{Sublattice-preserving symmetry}
\label{sec:two_sub_preserve}

We first consider the counterpart of Sec.~\ref{sec:one_sub_little} in which an arm-preserving spatial operation leaves each magnetic sublattice invariant. All quantities below are evaluated at a fixed generic incommensurate $\bQ$, with $R_g\bQ=\bQ$ exactly. In the position-based Fourier convention of Eq.~(\ref{eq:realspace_reconstruction}), the representation of $g=\{R_g|\boldsymbol\tau_g\}$ is 
\begin{equation}
 \mathcal U_g=e^{-i\bQ\cdot\boldsymbol\tau_g}
 \begin{pmatrix}\mathcal D_g&0\\0&\mathcal D_g\end{pmatrix}.
 \label{eq:preserve_rep}
\end{equation}
Consequently, Hamiltonian invariance requires
\begin{equation}
 \mathcal D_g\mathcal H_{\mu\mu}\mathcal D_g^\intercal =\mathcal H_{\mu\mu},
 \qquad \mathcal D_g \mathcal H_{12} \mathcal D_g^\intercal=\mathcal H_{12}.
 \label{eq:preserve_blocks}
\end{equation}
Unlike an exchanging operation, this symmetry does not relate $\mathcal H_{11}$ to $\mathcal H_{22}$ and does not require $\mathcal H_{12}$ to be Hermitian. Each diagonal block obeys the one-sublattice constraints separately, whereas the off-diagonal block belongs to the same invariant matrix space.

For a mirror or twofold rotation, as introduced in Eq. (\ref{eq:one_involution_principal}) we choose an orthonormal basis $(\vu a,\vu b,\vu c)$ such that
$\mathcal D_g=\mathrm{diag}(1,-1,-1)$ and $\vu b\cross\vu c=\vu a$. In this choice, the DM vectors are $\vb D^{(\mu)}=d^{(\mu)} \vu a$. The corresponding axes are listed in Table~\ref{tab:one_little_three}. The general allowed blocks are
\begin{equation}
\begin{aligned}
 \mathcal H_{\mu\mu}&= \mqty(
  h_{a}^{(\mu)}&0&0\\
 0&h_{b}^{(\mu)} &v^{(\mu)}-id^{(\mu)}\\
 0&v^{(\mu)}+id^{(\mu)}&h_{c}^{(\mu)}) \\
 \mathcal H_{12}&= \mqty(
 x_a&0&0\\0&x_{bb}&x_{bc}\\0&x_{cb}&x_{cc}). 
 \label{eq:preserve_model}
\end{aligned}
\end{equation}
Here $h_{a}^{(\mu)},h_{b}^{(\mu)},h_{c}^{(\mu)},v^{(\mu)},d^{(\mu)}$ are real, and the five entries of $\mathcal H_{12}$ are arbitrary complex numbers. In general, the two real transverse blocks in $\mathcal H_{11}$ and $\mathcal H_{22}$ cannot be diagonalized simultaneously by the same rotation of $(\vu b,\vu c)$.

The Hamiltonian separates into longitudinal and transverse sectors, $\mathcal H_{\text{2-sub}}\simeq L\oplus T$, where
\begin{equation}
 L=\mqty(h_{a}^{(1)}&x_a\\x_a^*&h_{a}^{(2)}),\qquad
 T=\begin{pmatrix}T_1&X_\perp\\X_\perp^\dagger&T_2\end{pmatrix}.
 \label{eq:preserve_sectors}
\end{equation}
The matrices $T_\mu$ and $X_\perp$ are the lower $2\times2$ blocks of Eq.~(\ref{eq:preserve_model}). Thus the linear spectrum separates into dimensions two and four, rather than the two three-dimensional sectors generated by sublattice exchange. The longitudinal sector alone cannot describe a circular hard-spin state because an amplitude proportional to a single real axis cannot satisfy both constraints in Eq.~(\ref{eq:hard_complex}). A state restricted to the transverse sector has $\vu n_\mu=\pm\vu a$. Although the symmetry separates the Hamiltonian into two sectors, it does not force a circular spiral to lie entirely in either sector. The optimal circular single-$Q$ minimization also admits amplitudes with both longitudinal and transverse components, whose tilted rotation planes may yield a lower energy.

We first derive exact circular solutions restricted to the transverse sector, for which the spiral normals are parallel or antiparallel to the symmetry axis. We then solve an isotropic symmetric-exchange model that illustrates how a lower-energy state can develop tilted normals by mixing longitudinal and transverse components. The extension to general anisotropy and simultaneous operations is discussed in Appendix~\ref{app:anisotropy_details}.

\subsubsection{Exact circular solutions in the transverse sector}
We first minimize within the transverse sector, taking $\vu n_\mu=\sigma_\mu\vu a$ with $\sigma_\mu=\pm1$. These states transform with a common character $-e^{-i\bQ\cdot\boldsymbol\tau_g}$ and preserve the operation up to a modulation phase. They preserve the ordinary spatial operation only when that character equals unity. Their energies must be compared with states outside this sector before identifying the constrained minimum. 

Define circular states for two sublattices as 
\begin{equation}
 \psi^{(\mu)}=e^{-i\phi_\mu}  p_{\sigma_\mu} \qquad \text{with } p_\sigma = \frac{1}{\sqrt2} \left( \vu b-i\sigma \vu c \right) 
 \label{eq:preserve_circular}
\end{equation}
Substitution into the energy gives 
 \begin{equation}
\begin{aligned}
E^{\text{intra}}_{\text{2-sub}} & = \sum_\mu \left( \bar r^{(\mu )} - \sigma_\mu d^{(\mu)}  \right), \\
\mathcal Z_{\sigma_1\sigma_2} &=p_{\sigma_1}^\dagger Xp_{\sigma_2} 
=\tfrac12\left[x_{bb}+\sigma_1\sigma_2x_{cc}
 +i\sigma_1x_{cb}-i\sigma_2x_{bc} \right],
 \label{eq:preserve_overlap}
\end{aligned}
\end{equation}
where $\bar r^{(\mu )}=(h_{b}^{(\mu )}+h_{c}^{(\mu )})/2$. 
Using the same phase convention as Eq.~(\ref{eq:phase_solution_general}), the exact branch minimum is
\begin{align}
 E_{\sigma_1\sigma_2}^{\rm min}
 &=\bar r^{(1)}+\bar r^{(2)}-\sigma_1 d^{(1)}-\sigma_2 d^{(2)}
 -2\abs{\mathcal Z_{\sigma_1\sigma_2}},
 \label{eq:preserve_branch_energy}
\end{align}
Then $ \Delta \phi =\arg \mathcal Z_{\sigma_1\sigma_2}+\pi\pmod{2\pi}$; if $\mathcal Z_{\sigma_1\sigma_2}=0$, the relative phase is arbitrary. 
The lowest transverse-sector energy is obtained by comparing these four explicit values ($\sigma_{1,2}=\pm1$); no scalar root search is needed. (The quantities $h_{a}^{(\mu)}$, $v^{(\mu)}$, and $x_a$ do not enter these branch energies, but they can determine stability against tilting the normals away from $\vu a$.)

\begin{table*}[tb]
\caption{Exact minimum of Eq.~(\ref{eq:preserve_isotropic}) for isotropic symmetric exchange when $d_1=p>0$, $d_2=-r<0$; $K_< =\frac{pr}{p+r}$ and $K_> =\frac{pr}{|p-r|}$. \footnote{For $d_1 d_2 >0$, $\vu n_1=\vu n_2=\mathrm{sgn}(d_1)\vu a$ and $E^{\text{min}}=J_1+J_2-2K-\abs{d_1}-\abs{d_2}$ is shown in Eq. (\ref{eq:EJ1J2pos}).}}
\label{tab:preserve_phase_diagram}
\begin{ruledtabular}
\begin{tabular}{lll}
Coupling regime & Minimizing normals & $E^{\min}$\\
\hline
$0<K\leq K_<$ & $\vu n_1=\vu a$, $\vu n_2=-\vu a$ & $J_1+J_2-p-r$\\
$K_<<K<K_>$ & Eq.~(\ref{eq:preserve_tilted_normals}) & $J_1+J_2-K-\tfrac12\big(pK/r+pr/K+rK/p\big)$\\
$K\geq K_>$, $p\ne r$ & $\vu n_1=\vu n_2=\mathrm{sgn}(p-r)\vu a$ & $J_1+J_2-2K-|p-r|$\\
\end{tabular}
\end{ruledtabular}
\end{table*}

\subsubsection{Isotropic symmetric exchange: an exact tilted solution}
An exactly solvable model exposes the competition between the two sublattice chiral preferences:
\begin{equation}
 \mathcal H_{\mu\mu}=J_{\mu} \mathbb I+i d_{\mu} [\vu a]_{\cross},
 \qquad \mathcal H_{12} =\mathcal J\mathbb I,
 \label{eq:preserve_isotropic}
\end{equation}
where $J_\mu,d_\mu$ are real and $\mathcal J$ may be complex. Define $K=\abs{\mathcal J}>0$. Normalized circular amplitudes $p(\vu n_1)$ and $p(\vu n_2)$ obey 
\begin{equation}
 |p(\vu n_1)^\dagger p(\vu n_2)|
 =\frac{1+\vu n_1\cdot\vu n_2}{2}.
 \label{eq:preserve_overlap_identity}
\end{equation}
Consequently, exact minimization over the relative phase gives
\begin{equation}
 E(\vu n_1, \vu n_2) =J_1+J_2-K
 -d_1\vu a\cdot\vu n_1-d_2\vu a\cdot\vu n_2
 -K\vu n_1\cdot\vu n_2.
 \label{eq:preserve_normal_energy}
\end{equation}
If $d_1d_2>0$, their common preferred normal $\vu n_1=\vu n_2=\mathrm{sgn}(d_1)\vu a$ minimizes every term simultaneously, giving the minimum
\begin{equation}
	E^{\text{min}}=J_1+J_2-2K-\abs{d_1}-\abs{d_2}. 
	\label{eq:EJ1J2pos}
\end{equation}

For opposite DM fields, the intra-sublattice DM interactions compete with the inter-sublattice symmetric exchange. Without loss of generality, $d_1=p>0$ and $d_2=-r<0$ are taken. At fixed $\vu n_2$, the optimal $\vu n_1$ is along
\begin{equation}
\begin{aligned}
 \vu n_1&=\frac{p\vu a+K\vu n_2}{\ell(x)}, \\
 \ell(x)&=\sqrt{p^2+K^2+2pKx},
\end{aligned}
\end{equation}
where $x=\vu a\cdot\vu n_2\in[-1,1]$. The remaining energy is
\begin{align}
 E(x)&=J_1+J_2-K+rx-\ell(x).
 \label{eq:preserve_scalar_min}
\end{align}
For $-1<x<1$, $E''(x)=(pK)^2/\ell(x)^3>0$, so the minimum is either the unique interior stationary point or an endpoint. The interior solution is
\begin{equation}
 x_* =\frac{(pK/r)^2-p^2-K^2}{2pK},
 \label{eq:preserve_scalar_solution}
\end{equation}
where $E(x_*) = J_1+J_2-K - pK /r + rx_*$. The endpoint energies are $E(x=\pm1) = J_1+J_2-K \pm r - \abs{p \pm K}$.

For $-1<x_*<1$, choose any unit vector $\vu u\perp\vu a$ and set
\begin{equation}
 \vu n_2=x_*\vu a+\sqrt{1-x_*^2}\vu u,\qquad
 \vu n_1=\frac{p\vu a+K\vu n_2}{pK/r}.
 \label{eq:preserve_tilted_normals}
\end{equation}
These formulas determine both spiral planes; their transverse frames and optimal relative phase follow from Eqs.~(\ref{eq:psi}) and (\ref{eq:phase_solution_general}). 
The condition $|x_*|\leq1$ is equivalent to $(p-K)^2\leq(pK/r)^2\leq(p+K)^2$. It holds for $K_<\leq K\leq K_>$, where $K_< =\frac{pr}{p+r}$ and $K_> =\frac{pr}{|p-r|}$ with $K_>=\infty$ when $p=r$. 
When $K$ is outside this interval, the minimum occurs at $x=\pm1$. The exact energies and normal configurations are summarized in Table~\ref{tab:preserve_phase_diagram}. Thus isotropic symmetric exchange does not guarantee an aligned-axis minimum when the sublattices are preserved rather than exchanged by the symmetry.

\subsection{Sublattice-exchange symmetry}
\label{sec:two_sub_exchange}

We now restore the wave-vector argument and consider a generic
incommensurate $\bQ=(q,q,0)$.  Suppose that a unitary space-group operation
$g=\{R_g| \boldsymbol\tau_g\}$ exchanges the two magnetic sublattices and
satisfies $R_g\bQ=\bQ$. The operation $g$ may be a mirror or glide, or a twofold rotation or screw rotation. The representation in the position-based gauge is
\begin{equation}
 \mathcal U_g =e^{-i\bQ\cdot\boldsymbol\tau_g}
 \mqty( 0&\mathcal{D}_g\\\mathcal{D}_g&0),
 \label{eq:exchange_position_rep}
\end{equation}
where $\mathcal{D}_g$ is an axial spin matrix as explained in Sec.~\ref{sec:one_sub_little}.  
The off-diagonal blocks implement the sublattice exchange $1\leftrightarrow2$. 
Invariance of the Hamiltonian then gives
\begin{align}
\begin{split}
\mathcal H_{22}&=\mathcal{D}_g\mathcal H_{11}\mathcal{D}_g^\intercal, \\
 \mathcal H_{12}&=\mathcal{D}_g\mathcal H_{12}^{\dagger}\mathcal{D}_g^\intercal,
 \qquad \mathcal H_{21}=\mathcal H_{12}^{\dagger}.
\end{split}
  \label{eq:exchange_block_reduction}
\end{align}
All blocks in this discussion are evaluated at the same $\bQ$.
Writing $\mathcal H_{\mu\mu}=\mathcal R^{(\mu)}+i[\vb D ^{(\mu)}]_{\cross}$, we obtain
\begin{equation}
 \mathcal R^{(2)}=\mathcal{D}_g\mathcal R^{(1)}\mathcal{D}_g^\intercal,\qquad
 \vb D^{(2)}=\mathcal{D}_g\vb D^{(1)}.
 \label{eq:exchange_intra_reduction}
\end{equation}
Thus $h_i^{(2)}=h_i^{(1)}$, and the symmetric-part eigenvectors can be
chosen as $\vu z_i^{(2)}=\mathcal{D}_g\vu z_i^{(1)}$. An exchanging symmetry
relates the two diagonal blocks; it does not require either block to
be separately invariant under $\mathcal{D}_g$.

By noting Eq.~(\ref{eq:exchange_block_reduction}), we can choose the basis of $\widetilde{\Psi}=(\widetilde \psi^{(1)},\widetilde \psi^{(2)})^\intercal$, where $\widetilde \psi^{(1)}=\psi^{(1)}$ and $\widetilde \psi^{(2)}=\mathcal{D}_g\psi^{(2)}$, to change $\mathcal{H}_{\text{2-sub}}$ to  
\begin{equation}
 \widetilde{\mathcal H}_{\text{2-sub}}=
 \mqty(H_0&K_g\\K_g&H_0),
 \label{eq:H2_sub}
\end{equation}
where the intra- and inter-sublattice interactions are both Hermitian and are 
\begin{align}
 H_0 &= \mathcal H_{11}= \mathcal{R}  + i [\bD]_{\cross},\\
 K_g &= \mathcal H_{12}\mathcal{D}_g = \mathcal{K} + i [\vb{M}]_{\cross}.
 \label{eq:inter_Kg}
\end{align}
respectively. 

The transformed normals are $\tilde{\vb n}_1=\vu n_1$ and $\tilde{\vb n}_2=\mathcal D_g\vu n_2$. The inter-sublattice energy is obtained through Eq.~(\ref{eq:main_overlap_trace}) with $\mathcal H_{12}=K_g$. Its explicit tensor expansion is given in Appendix~\ref{app:projector_details}.

\subsubsection{Exact minimum for isotropic symmetric exchange}
\label{sec:isotropic_symmetric_exchange}
Suppose $\mathcal R=J\mathbb{I}$ and $\mathcal K=J'\mathbb{I}$ in this exchange-adapted basis. The two linear sectors have matrices
\begin{equation}
 H_s=(J+sJ')\mathbb{I}+i[\vb D+s\vb M]_{\cross},\qquad s=\pm1,
 \label{eq:main_exchange_sectors}
\end{equation}
whose lowest eigenvalues are $e_s=J+sJ'-|\vb D+s\vb M|$. Each sector admits a normalized circular lowest eigenvector $w_s$, including an arbitrary circular choice if $\vb D+s\vb M=0$. Thus
\begin{equation}
 E_{\text{2-sub}}^{\min}=2\min_{s=\pm1}e_s
 \label{eq:main_exchange_minimum}
\end{equation}
is attained by
\begin{equation}
 \psi^{(1)}=w_s,\qquad \psi^{(2)}=s\mathcal D_gw_s.
 \label{eq:main_exchange_state}
\end{equation}
For a nonzero effective DM vector, both transformed normals align with $\vb D+s\vb M$. 

The two choices $s=\pm1$ correspond to the phase locking between $\psi^{(1)}$ and $\mathcal D_g \psi^{(2)}$ with relative phases of $0$ and $\pi$, respectively. Both choices therefore give the same relation between the spiral-plane normals, $\vu n_1= \mathcal D_g\vu n_2$. If $e_s<e_{-s}$, every minimizing state lies entirely in sector $s$ and satisfies $\psi^{(2)}=s\mathcal D_g\psi^{(1)}$. If $e_+=e_-$, either sector provides a minimizing circular spiral. Superpositions of minimizing modes from the two sectors also minimize the energy, provided they satisfy the normalization and circularity constraints on each sublattice. The proof is given in Appendix~\ref{app:exchange_proof}.

\subsubsection{Sublattice exchange sectors}

For general symmetric exchange interactions, each pure sublattice-exchange sector 
reduces to the one-sublattice problem with
$R_s=\mathcal R+s\mathcal K$ and
$\vb d_s=\vb D+s\vb M$, where $s=\pm1$.
The corresponding constrained energies follow from
the one-sublattice secular equation, including its
singular prescription, as detailed in
Appendix~\ref{sec:two_sub_sectors}.
However, sublattice-exchange symmetry does not generally
require the full constrained minimum to belong to
either pure sector.

To demonstrate this distinction, we introduce a
controlled symmetric anisotropy through the sector
Hamiltonians
\begin{equation}
    H_s(\eta)
    =(J+sJ')\mathbb I
    +\eta(\mathsf A+s\mathsf B)
    +i[\vb D+s\vb M]_{\cross},
    \qquad s=\pm1,
    \label{eq:anisotropic_sectors}
\end{equation}
where $\mathsf A$ and $\mathsf B$ are real symmetric matrices.

\begin{figure*}[t]
    \centering
    \includegraphics[width=0.90\textwidth]{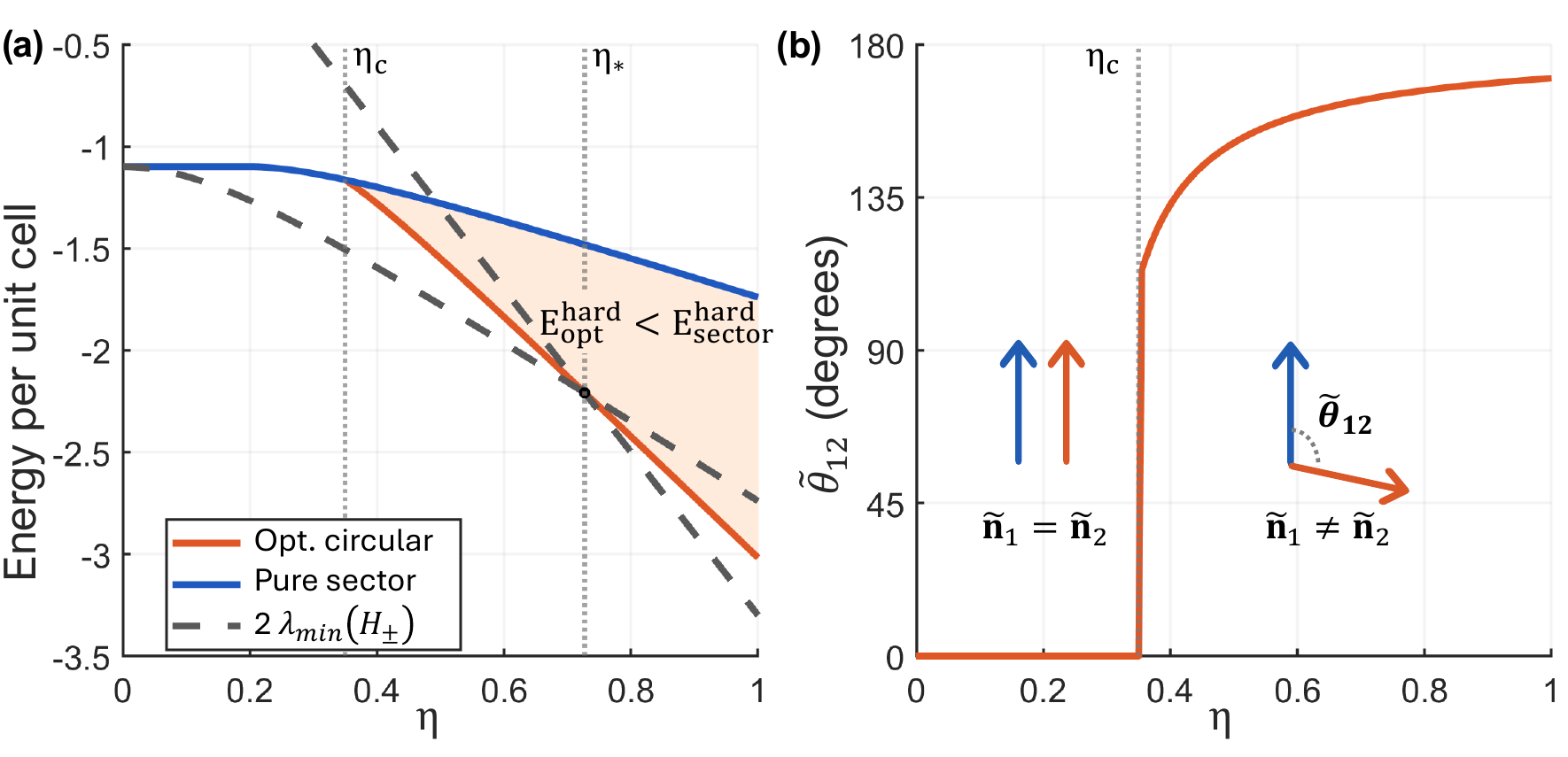}
    \caption{
    Anisotropy-induced departure from pure exchange
    sectors in a two-sublattice circular hard-spin model.
    (a) Lowest eigenvalue branches of the two exchange
    sectors, $2\lambda_{\min}(H_\pm)$ (gray dashed),
    optimal constrained circular single-$Q$ energy
    (orange), and lowest constrained pure-sector
    energy (blue), as functions of the symmetric
    anisotropy strength $\eta$.
    The lower envelope of the two eigenvalue branches
    gives the LT bound. The shaded region highlights
    the energy gain from allowing independent
    spiral-plane normals.
    The vertical dotted lines mark the departure
    from the pure-sector solution at
    $\eta_c\simeq0.35$ and the exchange-sector
    eigenvalue crossing at $\eta_*\simeq0.727$.
    At $\eta_*$, a sector-mixed circular state
    saturates the LT bound.
    (b) Relative angle
    $\widetilde{\theta}_{12}
    =\cos^{-1}(\widetilde{\vb n}_1
    \cdot\widetilde{\vb n}_2)$
    between the optimized normals in the
    exchange-adapted basis, where
    $\widetilde{\vb n}_1=\vb n_1$ and
    $\widetilde{\vb n}_2=\mathcal D_g\vb n_2$.
    The discontinuous change at $\eta_c$
    accompanies the departure from the
    pure-sector solution.
    The schematic arrows illustrate the
    relative normal orientations.
    Parameters are $J=0$, $J'=0.35$,
    $\vb D=0.30\hat{\vb e}_{\parallel}$,
    and $\vb M=0.10\hat{\vb e}_{\parallel}$,
    with $\mathsf A$ and $\mathsf B$ defined
    in Eq.~(\ref{eq:anisotropy_tensors}).
    All energies are per two-site unit cell
    in arbitrary units.
    }
\label{fig:sector_comparison}
\end{figure*}

We choose the anisotropy tensors in the crystallographic basis $(\hat{\vb e}_{\parallel},\hat{\vb e}_{\perp},\hat{\vb e}_z)$ of Eq.~(\ref{eq:diagonal_exchange_basis}) as
\begin{equation}
\mathsf A=\mathrm{diag}(-1,0,1),\qquad \mathsf B=\mathrm{diag}(-1,1,0),
\label{eq:anisotropy_tensors}
\end{equation}
with $\vb D=D_\parallel\hat{\vb e}_{\parallel}$ and $\vb M=M_\parallel\hat{\vb e}_{\parallel}$. This choice is compatible, at fixed $\bQ=(q,q,0)$, with an arm-preserving $C_{2[110]}$ operation that exchanges the sublattices and an arm-reversing $C_{2z}$ operation that preserves them. The corresponding Hamiltonian belongs to the $\mathcal T_{a-}$ symmetry class in Table~\ref{tab:app_two_forms}. The numerical parameters are specified in Fig.~\ref{fig:sector_comparison}.

Figure~\ref{fig:sector_comparison}(a) compares the
LT lower bound, the exact constrained pure-sector
energies, and the optimal circular hard-spin energy
obtained by optimizing two independent spiral-plane
normals. At small anisotropy, the optimal circular 
solution coincides with the lowest pure-sector branch.
Beyond $\eta_c\simeq0.35$, however, the optimal circular spiral loses the sublattice symmetry,
resulting in
\begin{equation}
    E_{\mathrm{LT}}
    \leq E_{\mathrm{opt}}^{\mathrm{hard}}
    < E_{\mathrm{sector}}^{\mathrm{hard}},
    \label{eq:anisotropy_energy_inequality}
\end{equation}
where
$E_{\mathrm{sector}}^{\mathrm{hard}}
=\min\{E_+^{\mathrm{hard}},E_-^{\mathrm{hard}}\}$.
The departure from the pure-sector solution is
accompanied by a discontinuous change in the relative
angle between the transformed spiral-plane normals,
as shown in Fig.~\ref{fig:sector_comparison}(b).

Thus, although sublattice-exchange symmetry block-diagonalizes the Hamiltonian, symmetric anisotropy can induce hybridization between sublattice-exchange sectors in the stabilized solution. This demonstrates the need for a two-normal optimization in the presence of general anisotropic interactions. Interestingly,
at a distinct anisotropy value
$\eta_*\simeq0.727$, the lowest eigenvalues
of the two exchange sectors become degenerate.
A suitable superposition of these modes
satisfies the local circularity constraints
and saturates the LT bound, as demonstrated
analytically in Appendix ~\ref{app:numerics}.

\section{Numerical application: wave-vector selection}
\label{sec:num_2sub}

\begin{figure*}[t]
\centering
\includegraphics[width=0.96\textwidth]{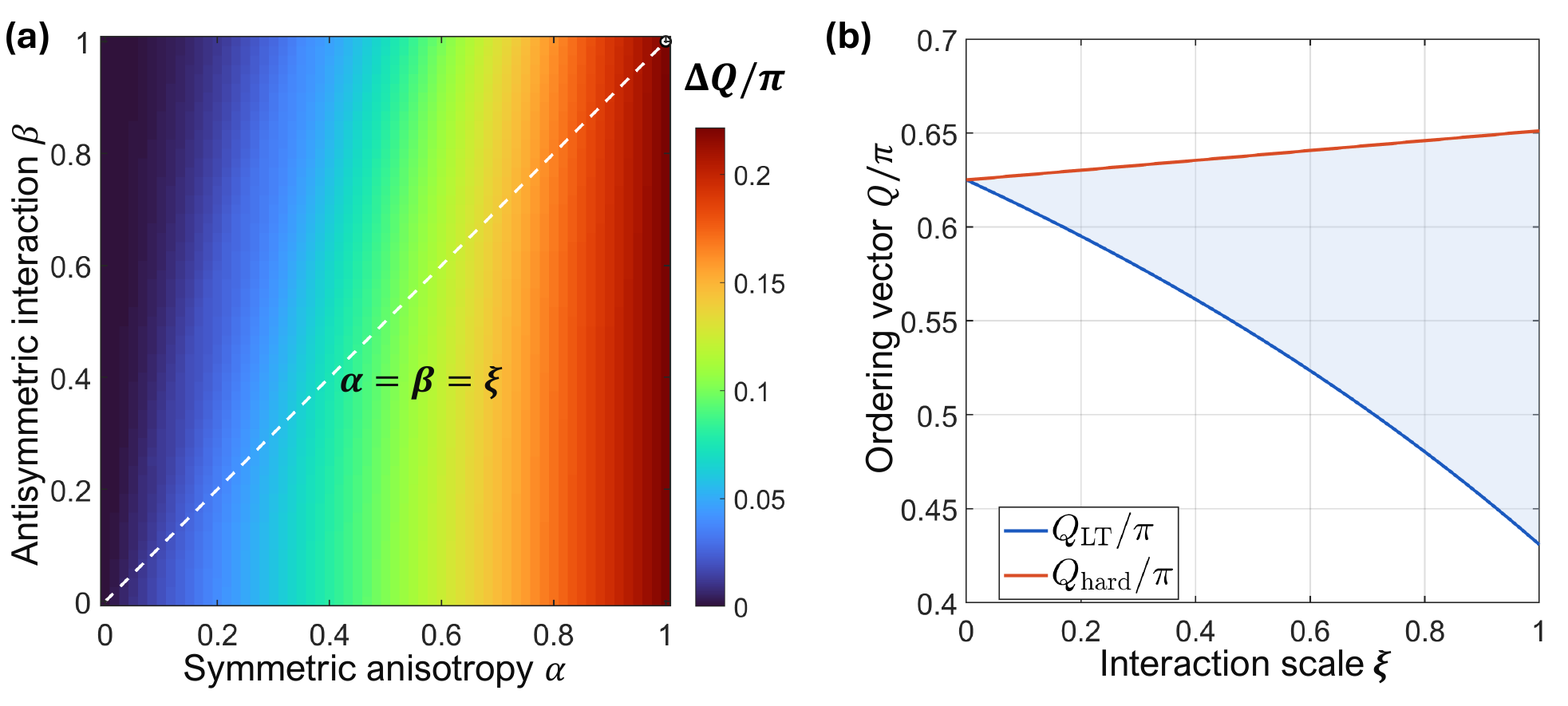}
\caption{Wave-vector selection under local circularity constraints in the exchange-symmetric two-sublattice model. (a) Difference $(Q_{\rm hard}-Q_{\rm LT})/\pi$ between the ordering wave vectors selected by the constrained circular energy and the LT lower bound, as a function of the symmetric anisotropy scale $\alpha$ and antisymmetric interaction scale $\beta$. The dashed line denotes the interpolation $\alpha=\beta=\xi$, and the marker indicates the benchmark $(\alpha,\beta)=(1,1)$. (b) Evolution of $Q_{\rm LT}/\pi$ and $Q_{\rm hard}/\pi$ along this interpolation. The wave vectors coincide in the isotropic limit and separate progressively as the non-isotropic interactions increase.}
\label{fig:LT_hardspin_comparison}
\end{figure*}

We now examine how local hard-spin constraints modify wave-vector selection in a two-sublattice model with sublattice-exchange symmetry. Unlike the fixed-$\mathbf Q$ analysis of Sec.~\ref{sec:two_sub_exchange}, the interaction coefficients here depend explicitly on the wave vector $q$. We consider a one-dimensional chain with two magnetic sites per unit cell, taking the lattice spacing $a=1$ and measuring energies in units of $J_1$. The symmetric anisotropic and antisymmetric interactions are scaled independently by two real dimensionless parameters, $\alpha$ and $\beta$, while the isotropic exchange is kept fixed.

In the exchange-adapted basis introduced in Sec.~\ref{sec:two_sub_exchange}, the Fourier-space interaction matrix takes the form
\begin{equation}
\widetilde{\mathcal H}(q;\alpha,\beta)=
\begin{pmatrix}
H_0(q;\alpha,\beta)&K_g(q;\alpha,\beta)\\
K_g(q;\alpha,\beta)&H_0(q;\alpha,\beta)
\end{pmatrix},
\label{eq:num_H}
\end{equation}
where
\begin{align}
H_0
&=2J_2\cos q\,\mathbb I
+\alpha K_{\rm an}\mathsf A
+i\beta d\sin q[\hat{\vb e}_{\parallel}]_{\cross},
\label{eq:num_Hmumu}\\
K_g
&=2J_1\cos(q/2)\mathbb I
+2\alpha c_0\cos(q/2)\mathsf B
\nonumber\\
&\quad+2i\beta b_0\sin(q/2)
[\hat{\vb e}_{\parallel}]_{\cross}.
\label{eq:num_H12}
\end{align}
The matrices $\mathsf A$ and $\mathsf B$ are the real symmetric anisotropy tensors defined in Eq.~(\ref{eq:anisotropy_tensors}), expressed in the basis $(\hat{\vb e}_{\parallel},\hat{\vb e}_{\perp},\hat{\vb e}_z)$. We choose $J_1=1$, $J_2=0.45$, $K_{\rm an}=0.60$, $d=0.20$, $b_0=0.25$, and $c_0=0.45$. The factors $\cos(q/2)$ and $\sin(q/2)$ describe symmetric and antisymmetric combinations of neighboring inter-sublattice bonds in the position-based Fourier gauge, with basis positions $r_1=0$ and $r_2=1/2$. In this representation, the physical inter-sublattice block is related to the exchange-adapted block by $\mathcal H_{12}=K_g\mathcal D_g$, where $\mathcal D_g=\mathrm{diag}(1,-1,-1)$. Both $H_0$ and $K_g$ are Hermitian, and the full matrix satisfies the Fourier-reality condition $\widetilde{\mathcal H}(-q;\alpha,\beta)=\widetilde{\mathcal H}(q;\alpha,\beta)^*$.

The structure of Eq.~(\ref{eq:num_H}) admits an exact exchange-sector decomposition. Introducing
\begin{equation}
H_s(q;\alpha,\beta)=H_0(q;\alpha,\beta)+sK_g(q;\alpha,\beta),\qquad s=\pm1,
\label{eq:num_sectors}
\end{equation}
a unitary transformation block-diagonalizes the full $6\times6$ interaction matrix into $H_+\oplus H_-$. Equivalently, the exchange operator $\mathcal X=\begin{pmatrix}0&\mathbb I\\ \mathbb I&0\end{pmatrix}$ commutes with $\widetilde{\mathcal H}$ at every $q$. A nondegenerate LT eigenvector therefore has definite exchange parity, $\widetilde v^{(2)}=s\widetilde v^{(1)}$, and consequently equal sublattice weights,
\begin{equation}
\widetilde v^{(1)\dagger}\widetilde v^{(1)}
=\widetilde v^{(2)\dagger}\widetilde v^{(2)}
=\frac12.
\label{eq:num_equal_weights}
\end{equation}
This property does not guarantee local circularity, since $\widetilde v^{(\mu)\intercal}\widetilde v^{(\mu)}$ need not vanish. The exchange-sector decomposition is exact for the linear problem, but the locally constrained minimum must still be determined within the circular two-sublattice manifold.

At fixed $q$, the LT lower bound is
\begin{equation}
E_{\rm LT}(q;\alpha,\beta)
=2\min_{s=\pm1}
\lambda_{\min}[H_s(q;\alpha,\beta)].
\label{eq:num_LT_energy}
\end{equation}
The corresponding circular hard-spin energy $E_{\rm hard}(q;\alpha,\beta)$ is obtained by minimizing the phase-reduced expression of Eq.~(\ref{eq:E2sub_final}) over two independent unit spiral-plane normals in the exchange-adapted basis. The ordering vectors $Q_{\rm LT}$ and $Q_{\rm hard}$ are then determined by minimizing the respective energies over $0\leq q\leq\pi$.

In the isotropic limit $\alpha=\beta=0$, the two energies coincide because the spin-degenerate lowest LT eigenspace admits locally normalized circular states. Their common dispersion is $E_{\rm LT}^{(0)}(q)=E_{\rm hard}^{(0)}(q)=4J_2\cos q-4J_1\cos(q/2)$, which is minimized at
\begin{equation}
Q_0=2\cos^{-1}\left(\frac{J_1}{4J_2}\right).
\label{eq:num_isotropic_Q}
\end{equation}
For the chosen exchange parameters, this gives $Q_0/\pi\simeq0.625$. Thus, before the anisotropic interactions are introduced, the LT and circular hard-spin formulations select the same ordering wave vector.

Figure~\ref{fig:LT_hardspin_comparison} shows the evolution of the wave-vector separation as the interactions are varied over $0\leq\alpha,\beta\leq1$. The left panel presents $(Q_{\rm hard}-Q_{\rm LT})/\pi$ in the two-parameter space, while the right panel follows the interpolation $\alpha=\beta=\xi$. The two wave vectors coincide in the isotropic limit and separate progressively as the non-isotropic interactions increase. For the present model, the separation depends predominantly on the symmetric anisotropy scale $\alpha$, with a weaker dependence on the antisymmetric interaction scale $\beta$.

At the benchmark point $(\alpha,\beta)=(1,1)$, the LT and constrained minima occur at $Q_{\rm LT}/\pi\simeq0.431$ and $Q_{\rm hard}/\pi\simeq0.651$, respectively, corresponding to a wave-vector separation of approximately $0.220\pi$. Their minimum energies per two-site unit cell are $E_{\rm LT}\simeq-4.155$ and $E_{\rm hard}\simeq-3.592$. The appreciable shift illustrates how imposing local hard-spin constraints can modify the ordering vector selected by the LT approximation.

The lowest LT eigenvectors at both minimizing wave vectors have equal sublattice weights but fail the local circularity test of Eq.~(\ref{eq:eigenvector_test}). The corresponding normalized circularity residuals are approximately $0.984$ at $Q_{\rm LT}$ and $0.906$ at $Q_{\rm hard}$, with identical values on the two sublattices. Since both lowest LT eigenvalues are nondegenerate, the LT eigenvectors cannot be combined with other degenerate lowest modes to restore circularity at these wave vectors. Thus, equal sublattice weights alone are insufficient to reach the LT bound.

Interestingly, the optimized hard-spin state at the benchmark point remains within the negative exchange sector. Its transformed spiral-plane normals coincide to numerical precision, $\widetilde{\vb n}_1=\widetilde{\vb n}_2$, and the relative phase is approximately $\pi$, giving $\widetilde\psi^{(2)}\simeq-\widetilde\psi^{(1)}$. In contrast to the sector-mixed solution discussed in Sec.~\ref{sec:two_sub_exchange}, the wave-vector shift in the present model does not require mixing the two exchange sectors.

The origin of this shift is the wave-vector dependence of the constraint penalty,
\begin{equation}
\Delta E_{\rm con}(q)
=E_{\rm hard}(q)-E_{\rm LT}(q)\geq0.
\label{eq:num_constraint_penalty}
\end{equation}
At the constrained minimum, this penalty is approximately $0.363$ per unit cell. The constrained correction therefore modifies the energy landscape rather than merely adding a constant offset to the LT dispersion. 
The present example demonstrates that local hard-spin constraints can substantially change wave-vector selection even when sublattice-exchange symmetry enforces equal LT weights and the optimized hard-spin state remains within a pure exchange sector. 
The corresponding energy dispersions are shown
in Fig.~\ref{fig:app_LT_dispersion}, while
additional numerical consistency checks are
presented in Appendix~\ref{app:two_sub_validation}.

\section{Discussion and conclusion}
\label{sec:conclusion}

We have developed a constrained formulation for classical circular single-$Q$ spirals with one or two magnetic sublattices. The central distinction from the Luttinger--Tisza (LT) approach is that equal sublattice weights alone do not guarantee local spin normalization: each sublattice amplitude must also satisfy the circularity condition. For one sublattice, this constraint reduces the minimization to a secular equation for the spiral-plane normal, including a singular branch that must be retained for the general solution. For two sublattices, exact elimination of the relative in-plane phase reduces the problem to an optimization over two unit normals. The projector representation of the inter-sublattice overlap makes this reduction independent of the choice of transverse frames and remains regular for parallel and antiparallel normals.

Crystal symmetry provides further restrictions on the interaction matrices and, in selected cases, permits exact constrained solutions. By distinguishing operations that preserve or reverse the Fourier arm from those that preserve or exchange sublattices, we obtain the allowed Hamiltonian structures without imposing the symmetry of the parent crystal on the ordered spin configuration. The resulting classification also incorporates the compatibility conditions among sublattice permutations imposed by the parent group. Within these symmetry settings, competing intra-sublattice Dzyaloshinskii--Moriya interactions can stabilize spirals with tilted relative planes, even for isotropic symmetric exchange. Conversely, sublattice-exchange symmetry with isotropic symmetric interactions in the exchange-adapted basis admits an exact circular state that saturates the LT bound at fixed $\mathbf Q$.

The anisotropic exchange-sector example reveals an important limitation of this simplification. Although sublattice-exchange symmetry block-diagonalizes the linear Hamiltonian, it does not generally restrict the constrained minimum to either pure exchange sector. For the model considered here, numerical optimization over independent spiral-plane normals yields a lower energy than the constrained pure-sector solutions beyond a parameter-dependent anisotropy threshold. The accompanying reorientation of the transformed normals illustrates how symmetric anisotropy can favor a sector-mixed circular state. Thus, symmetry-sector decomposition simplifies the spectral problem but does not, by itself, resolve the local hard-spin constraints.

A separate anisotropic two-sublattice benchmark demonstrates that the consequences of these constraints extend beyond spiral-plane orientation. An exact algebraic relation enforces equal sublattice weights in nondegenerate LT eigenmodes, but does not guarantee local circularity. By independently varying symmetric anisotropy and antisymmetric exchange, we show that the LT and constrained ordering vectors coincide in the isotropic limit and progressively separate as non-isotropic interactions are introduced. For the chosen benchmark, the wave-vector difference reaches approximately $\Delta Q/\pi=0.220$. The two-parameter scan further demonstrates that this discrepancy persists over a finite region of interaction strengths and is driven predominantly by symmetric anisotropy in the present model. Thus, the local spin constraints can influence wave-vector selection even when the usual equal-sublattice-weight condition is satisfied.

The present results concern classical circular single-$Q$ states at generic incommensurate wave vectors. Except where a constrained configuration saturates the global LT bound, minimization within this manifold does not establish the unrestricted magnetic ground state. States involving additional Fourier components, higher harmonics, or noncoplanar order may compete energetically, while commensurate wave vectors can admit local spin configurations that do not satisfy the generic circularity criterion. Nevertheless, the analytical reductions and symmetry constraints provide a systematic framework for constructing and optimizing physically admissible single-$Q$ reference states and for comparing them with more general magnetic textures.

\section*{Acknowledgments}
This work was supported by National Science and Technology Council (NSTC) of Taiwan under Grant No. NSTC 115-2112-M-110-023. 

\section*{Data Availability}

The data that support the findings of this study are available
from the corresponding author upon reasonable request.

\clearpage
\appendix
\section{Derivations for two-sublattice systems}
\label{app:technical}

\subsection{Explicit transverse frames and component overlaps}
\label{app:frames}
The phase-independent expression in Eq.~(\ref{eq:main_overlap_trace}) is valid for arbitrary inter-sublattice matrices without confronting the choice of the transverse frame. Here we provide explicit formulas for a restricted but useful family of interactions.
For the explicit reductions below we parameterize the inter-sublattice block as
\begin{align}
        \mathcal{H}_{12} &= \tilde{A}\mathbb{I} + \tilde{B} \vu{t} \vu{t}^{\intercal} + i \tilde{C} [\vu{M}]_{\cross}, 
     \label{eq:inter_gen}
\end{align}
where $\tilde A$, $\tilde B$, and $\tilde C$ are complex and $\vu t$ and $\vu M$ are real unit vectors. This parameterization contains one symmetric anisotropy axis and one antisymmetric direction; it is not the most general complex inter-sublattice matrix. In particular, the real and imaginary symmetric parts need not share principal axes. The formulation of Sec.~\ref{sec:exact} and Eq.~(\ref{eq:Emin_Z}) apply to an arbitrary inter-sublattice block.

The energy $E_{\text{2-sub}}'=\Psi^\dagger\mathcal H_{\text{2-sub}}\Psi +\text{(constraints)}$ splits into intra- and inter-sublattice parts. With $\vu{n}_{\mu}=\sum_i\alpha_i^{(\mu)}\vu{z}_i^{(\mu)}$ and $\sum_i(\alpha_i^{(\mu)})^2=1$, the intra-sublattice part reads

\begin{equation}
    \begin{split}
        E^{\text{intra}}_{\text{2-sub}} &=\sum_{\mu}\sum_{i} \left[ \frac{1}{2} h_i^{(\mu)} \left(1-\left(\alpha_i^{(\mu)}\right) ^2 \right) - D_i^{(\mu)} \alpha_i^{(\mu)}\right] \\
        & \qquad + \frac{1}{2} \sum_{\mu}\lambda^{(\mu)} \left(\sum_{i} \left(\alpha_i^{(\mu)}\right) ^2 -1 \right) , \label{intra_2sub}
    \end{split}
\end{equation}
exactly as in Eq.~(\ref{eq:E_1_lagrange}) for each sublattice, while the inter-sublattice interaction energy  
\begin{equation}
    \begin{split}
        E_{\text{2-sub}}^{\text{inter}} &= \psi^{(1)\dagger} \mathcal{H}_{12} \psi^{(2)} + \psi^{(2)\dagger} \mathcal{H}_{21} \psi^{(1)} \\      
        &= \Re(\tilde{A}) \left[ \vu{s}_{1}^{(1)} \cdot \vu{s}_{1}^{(2)} +\vu{s}_{2}^{(1)} \cdot \vu{s}_{2}^{(2)} \right] \\
        &+ \Im(\tilde{A}) \left[ \vu{s}_{1}^{(1)} \cdot \vu{s}_{2}^{(2)} - \vu{s}_{2}^{(1)} \cdot \vu{s}_{1}^{(2)} \right] \\
        &+ \Re(\tilde{B}) \left[ (\vu{t} \cdot \vu{s}_{1}^{(1)})(\vu{t} \cdot \vu{s}_{1}^{(2)}) + (\vu{t} \cdot \vu{s}_{2}^{(1)})(\vu{t} \cdot \vu{s}_{2}^{(2)})\right] \\
        &+ \Im(\tilde{B}) \left[ (\vu{t} \cdot \vu{s}_{1}^{(1)})(\vu{t} \cdot \vu{s}_{2}^{(2)}) - (\vu{t} \cdot \vu{s}_{2}^{(1)})(\vu{t} \cdot \vu{s}_{1}^{(2)})\right] \\
        &- \Re(\tilde{C}) \left[ \vu{M} \cdot \left(\vu{s}_{1}^{(1)} \cross \vu{s}_{2}^{(2)} - \vu{s}_{2}^{(1)} \cross \vu{s}_{1}^{(2)} \right) \right] \\
        &+ \Im(\tilde{C}) \left[ \vu{M} \cdot \left( \vu{s}_{1}^{(1)} \cross \vu{s}_{1}^{(2)} +  \vu{s}_{2}^{(1)} \cross \vu{s}_{2}^{(2)} \right)\right] .
        \label{eq:inter_1}
    \end{split}
\end{equation}

\begin{figure}[t]
    \centering
    \includegraphics[width=0.98\columnwidth]{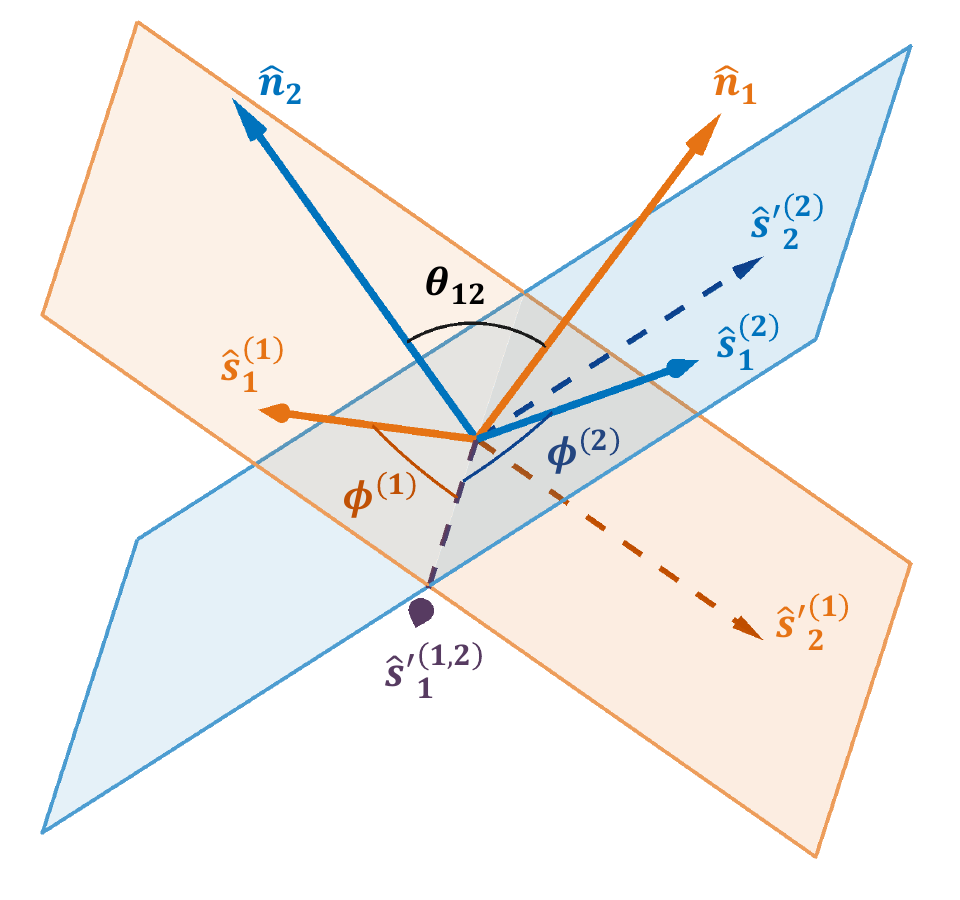}
    \caption{
    Geometry of the two coupled spiral frames.
    Each sublattice $\mu=1,2$ has an orthonormal frame
    $\{\vu s_1^{(\mu)},\vu s_2^{(\mu)},\vu n_\mu\}$,
    with $\vu n_\mu$ the oriented spiral-plane normal.
    For nonparallel and nonantiparallel normals,
    the common reference axis is
    $\vu s_1^{\prime(1)}=\vu s_1^{\prime(2)}$.
    The remaining reference axes satisfy
    $\vu s_2^{\prime(\mu)}
    =\vu n_\mu\cross\vu s_1^{\prime(\mu)}$.
    In-plane rotations by $\phi^{(\mu)}$ relate the
    reference frames to the physical spiral frames.
    The angle between the normals is
    $\theta_{12}=\cos^{-1}(\vu n_1\cdot\vu n_2)$.
    }
    \label{fig:spiral_frames}
\end{figure}

For nonparallel and nonantiparallel normals, the two
spiral planes intersect along a common axis.
The geometry of the corresponding reference frames
is illustrated in Fig.~\ref{fig:spiral_frames}.
A convenient choice is

\begin{equation}
\begin{split}
\vu{s}_1^{\prime(1)}
&=\vu{s}_1^{\prime(2)}
=\frac{\vu n_1\cross\vu n_2}{s},\\
\vu{s}_2^{\prime(1)}
&=\frac{c\vu n_1-\vu n_2}{s},\\
\vu{s}_2^{\prime(2)}
&=\frac{\vu n_1-c\vu n_2}{s},
\end{split}
\label{eq:primed_frame}
\end{equation}
where 
\begin{equation}
	c =\vu{n}_1 \cdot \vu{n}_2 =\cos\theta,\qquad
s=\| \vu{n}_1 \cross \vu{n}_2 \|=\sin \theta>0.
\end{equation}
This choice of the primed frame is geometrically explained as follows. The overlap matrix in the unprimed frame
\begin{equation}
    G= \mqty( 
        \vu{s}_1^{(1)}\vdot\vu{s}_1^{(2)}        & \vu{s}_1^{(1)}\vdot\vu{s}_2^{(2)}\\
        \vu{s}_2^{(1)}\vdot\vu{s}_1^{(2)}        & \vu{s}_2^{(1)}\vdot\vu{s}_2^{(2)}   ),
    \label{eq:frame_overlap}
\end{equation}
and that in the primed frame 
\begin{equation}
    G'=  \mqty(\vu{s}_1^{\prime (1)}\vdot\vu{s}_1^{\prime (2)} &  \vu{s}_1^{\prime(1)}\vdot\vu{s}_2^{\prime(2)} \\ 
    \vu{s}_2^{\prime (1)}\vdot\vu{s}_1^{\prime (2)}        & \vu{s}_2^{\prime (1)}\vdot\vu{s}_2^{\prime (2)}  )    = 
     \mqty(1  &  0 \\  0       &  \cos \theta ) 
\end{equation}
are related by $G'=U_1^\intercal G U_2$, where 
\begin{equation}
    U_\mu= \mqty(
        \cos\phi^{(\mu)} & -\sin\phi^{(\mu)}\\
        \sin\phi^{(\mu)} & \cos\phi^{(\mu)} ).
 \end{equation}
The diagonal form $G'=\mathrm{diag}(1,c)$ is a signed singular-value form of $G$. Its singular values are $1$ and $|c|$. When $c<0$, converting this signed form to a conventional nonnegative singular-value decomposition requires a reflection in one of the two-dimensional bases, rather than an in-plane phase rotation.

In these reference frames, the inter-sublattice overlap is
\begin{equation}
\begin{split}
 \mathcal{Z}  & \left(\vu{n}_1 , \vu{n}_2  \right) = \frac{1}{2} \left\{ \tilde{A} (1+c) - \tilde{C} \left[ (m_1 + m_2) + i m_3 \right] \right. \\
&\left. + \tilde{B} \left[
\frac{t_3^2- (t_1 -c t_2) (t_2 -c t_1)}{1-c^2} -i 
\frac{t_3 (t_1+t_2)}{1+c}\right] \right\},
\end{split}
\label{eq:Z_explicit}
\end{equation}
where
\begin{equation}
\begin{aligned}
 t_\mu&=\vu t\cdot\vu n_{\mu},&m_\mu&=\vu M\cdot\vu n_{\mu },\quad\mu=1,2,\\
 t_3&=\vu t\cdot(\vu n_1 \cross\vu n_2),&
 m_3&=\vu M\cdot(\vu n_1 \cross\vu n_2).
\end{aligned}
\label{eq:tandm}
\end{equation}
One can observe that once anisotropy is considered ($\tilde{B} $), the problem becomes complex (even contains $\vu{n}_1 \cdot \vu{n}_2 $ in the denominator). 
The complex interaction coefficients ($\tilde{A},\tilde{B},\tilde{C}$) further complicate the analytical determination of the phase of \(\mathcal Z\). 

For parallel or antiparallel normals ($\vu{n}_1 = \pm \vu{n}_2$), the intersection-based frames are
undefined. In these cases one should choose any regular orthonormal
transverse frames and evaluate the overlap directly from Eq.~(\ref{Z}),
rather than use the singular coordinate expression
(\ref{eq:Z_explicit}). 
The phase-minimized energy remains frame independent.

To impose their normalization during differentiation, introduce the Lagrangian
$E_{\text{2-sub}}'=E_{\text{2-sub}}+\tfrac12\sum_\mu\lambda^{(\mu)}(\vu n_\mu^\intercal\vu n_\mu-1)$. 
Differentiating this Lagrangian gives the exact stationarity
conditions (for \(\mathcal Z\neq0\))
\begin{equation}
    \left(\lambda^{(\mu)}-h_i^{(\mu)}\right)\alpha_i^{(\mu)}
    = D_i^{(\mu)} + \Gamma_i^{(\mu)},
    \qquad \mu=1,2,
    \label{eq:alpha_stationary_general}
\end{equation}
with the effective inter-sublattice field
\begin{equation}
    \Gamma_i^{(\mu)}
    = \frac{2}{\abs{\mathcal Z}}\,
    \Re\!\left(\mathcal Z^{*}\,
    \frac{\partial \mathcal Z}{\partial \alpha_i^{(\mu)}}\right).
    \label{eq:Gamma_def}
\end{equation}
Equation~(\ref{eq:alpha_stationary_general}) has the same structure as
the one-sublattice equation (\ref{eq:ai}) with \(D_i\to D_i+\Gamma_i\).  Consequently, the constraint closes into a secular equation for each sublattice,
\begin{equation}
    \sum_{i}\left(\frac{D_i^{(\mu)}+\Gamma_i^{(\mu)}}
    {\lambda^{(\mu)}-h_i^{(\mu)}}\right)^{2}=1,
    \qquad \mu=1,2,
    \label{eq:lambda_mu}
\end{equation}
These are implicit stationarity relations because $\Gamma^{(\mu)}$ depends on both sublattices, $\alpha^{(1,2)}$. They do not define independent scalar root searches or inherit the one-sublattice largest-root selection rule. 

\subsection{Energy formulation in the sublattice-exchange system}
\label{app:projector_details}
In Sec.~\ref{sec:two_sub_exchange}, we observed that in the presence of sublattice-exchange symmetry $\mathcal U_g : 1\leftrightarrow2$, the tilde frame 
\begin{equation}
	\widetilde \psi^{(1)}=\psi^{(1)}, \quad \widetilde \psi^{(2)}=\mathcal{D}_g\psi^{(2)}
\end{equation}
can simplify the Hamiltonian: both sublattice spins have the identical intra-sublattice interaction $H_0$, and the inter-sublattice interaction becomes Hermitian $K_g=K_g^\dagger$. 
In this Appendix, we will introduce a projector formalism to write the energy of the two-sublattice system as a function of normals, $\vb{n}_{1,2}$. 

The intra-sublattice interaction energy, 
\begin{equation}
E^{\text{intra}}_{\text{2-sub}} = \sum_{\mu =1,2}\widetilde \psi^{(\mu)\dagger} H_0 \widetilde \psi^{(\mu)} ,
\end{equation}
was evaluated in Sec.~\ref{sec:one_sub_theory}. 
The intra-sublattice contributions have identical functional forms in the exchange-adapted basis.
However, this does not imply that the minimizing amplitudes on the two sublattices must be identical, since the inter-sublattice interaction can favor a configuration outside the pure exchange sectors.
The inter-sublattice energy is similar to that derived in Sec.~\ref{sec:two_sub}, giving  
\begin{equation}
\begin{split}
 E^{\text{inter}}_{\text{2-sub}}  & =e^{-i\Delta\phi} \widetilde \psi^{(1)\dagger} K_g \widetilde\psi^{(2)} + \mathrm{c.c.} \\
 & = - 2 \abs{\mathcal Z},
 \end{split} 
 \label{E_inter_subsym}
\end{equation}
after optimizing the phase $\Delta\phi$. 

Introducing the circular-polarization projectors 
\begin{equation}
 \Pi_\mu = \widetilde \psi^{(\mu)} \widetilde \psi^{(\mu)\dagger} = \frac{1}{2} \left( \mathbb{I} - \tilde{\vb{n}}_{\mu} \tilde{\vb{n}}_{\mu}^\intercal  - i [\tilde{\vb{n}}_{\mu}]_{\cross} \right),
\end{equation}
where $\tilde{\vb{n}}_\mu  = -i \widetilde \psi^{(\mu)} \cross \widetilde \psi^{(\mu)*}$, 
the squared modulus of the inter-sublattice overlap can be written as
\begin{equation}
\begin{aligned}
\abs{\mathcal Z \left(\tilde{\vb{n}}_1,\tilde{\vb{n}}_2 \right)}^2
&=(\widetilde \psi^{(1) \dagger} K_{g} \widetilde \psi^{(2)})
  (\widetilde \psi^{(2)\dagger} K_{g} \widetilde \psi^{(1)})\\
&=\trace 
  \left( \Pi_1K_g \Pi_2 K_g \right) \\
  & =\frac14\left(F_K+F_M+F_{KM}\right),
\end{aligned}
\end{equation}
where 
\begin{equation}
\begin{aligned}
F_K= &\trace (\mathcal{K} ^2)
-\tilde{\vb{n}}_{1} ^{\intercal} \mathcal{K} ^2 \tilde{\vb{n}}_{1} -\tilde{\vb{n}}_{2}^{\intercal}\mathcal{K} ^2 \tilde{\vb{n}}_{2} \\
& +(\tilde{\vb{n}}_{1}^{\intercal}\mathcal{K} \tilde{\vb{n}}_{2})^2
+2\tilde{\vb{n}}_{1}^{\intercal}\mathrm{cof}(\mathcal{K} )\tilde{\vb{n}}_{2}, \\ 
F_M= &\big[\vb{M}\cdot(\tilde{\vb{n}}_{1} +\tilde{\vb{n}}_{2})\big]^2
+\big[\vb{M} \cdot(\tilde{\vb{n}}_{1} \cross \tilde{\vb{n}}_{2})\big]^2,\\ 
F_{KM}= &2 \left[
(1-\tilde{\vb{n}}_{1} \cdot \tilde{\vb{n}}_{2})(\tilde{\vb{n}}_{1} +\tilde{\vb{n}}_{2})^{\intercal}\mathcal{K} \vb{M}\right.\\
&\quad\left.
-\trace(\mathcal K)\,\vb{M}\cdot(\tilde{\vb{n}}_{1}+\tilde{\vb{n}}_{2} ) \right.\\
&  \quad \left.
+(\vb{M}\cdot \tilde{\vb{n}}_{1})\,\tilde{\vb{n}}_{2}^{\intercal} \mathcal{K} \tilde{\vb{n}}_{2}
+(\vb{M} \cdot \tilde{\vb{n}}_{2})\,\tilde{\vb{n}}_{1}^{\intercal} \mathcal{K}  \tilde{\vb{n}}_{1}
\right].
\end{aligned}
\end{equation}
The cofactor matrix of $\mathcal{K}$ is
\begin{equation}
\mathrm{ cof }( \mathcal{K} )
=\mathcal{K} ^2-\trace (\mathcal{K} ) \mathcal K
+\frac12\left[(\trace \mathcal K)^2-\trace(\mathcal{K}^2)\right] \mathbb{I}.
\end{equation}
The derivation uses the following identities, with $K$ real symmetric:
\begin{equation}
\begin{aligned}
{}[\vb{x}]_{\cross} [\vb{y} ]_{\cross} &=\vb{y} \vb{x}^\intercal -(\vb{x} \cdot \vb{y}) \mathbb{I}, \\
\trace ([\vb{x}]_{\cross} K [\vb{y}]_{\cross} K)
&=-2 \vb{x}^\intercal \mathrm{cof}(K) \vb{y}.
\end{aligned}
\end{equation}
Stationary configurations follow by varying the energy
\begin{equation}
\begin{aligned}
E(\tilde{\vb{n}}_1,\tilde{\vb{n}}_2)&=\sum_{\mu=1}^2\left[\tfrac12\big(\trace \mathcal R-\tilde{\vb{n}}_\mu^\intercal \mathcal R \tilde{\vb{n}}_\mu\big)
 -\bD\cdot \tilde{\vb{n}}_\mu\right] \\
 &\quad-2 \abs{\mathcal Z \left(\tilde{\vb{n}}_1,\tilde{\vb{n}}_2 \right)}.
 \end{aligned}
\end{equation}
When $|\mathcal Z|>0$, tangent stationarity is
\begin{equation}
 (\mathbb I-\tilde{\vb{n}}_\mu \tilde{\vb{n}}_\mu^\intercal)
 \left(\mathcal R \tilde{\vb{n}}_\mu+\bD+
 \frac{\nabla_{\tilde{\vb{n}}_\mu}\abs{\mathcal Z}^2}{\abs{\mathcal Z}} \right)=0.
 \label{eq:exchange_tangent}
\end{equation}
These are coupled nonlinear equations on $S^2\times S^2$; they do not reduce to independent secular equations. 
These equations are equivalent to the stationarity conditions in Eqs.~(\ref{eq:lambda_mu}) and (\ref{eq:nonlinear_eigen}).

Note that the normals in the transformed basis obey $\tilde{\vb n}_1=\vu n_1$ and $\tilde{\vb n}_2=\mathcal D_g\vu n_2$.

\subsubsection{Proof of the isotropic exchange-sector minimum}
\label{app:exchange_proof}
This Appendix is to explain the exchange symmetric or antisymmetric in spins on two sublattices in Sec.~\ref{sec:isotropic_symmetric_exchange}. 
Continue the sublattice-exchange symmetric system in Appendix~\ref{app:projector_details}. 
We consider isotropic symmetric exchange in the tilde frame, 
\begin{equation}
	\mathcal{R}  = J  \mathbb{I} \quad \text{ and } \quad \mathcal{K}  = J'  \mathbb{I},
\end{equation}
in $H_0 = \mathcal{R}+ i [ \vb D ]_{\cross}$ and $K_g = \mathcal{K}+ i [ \vb M]_{\cross}$. 
In this case, the phase-minimized inter-sublattice contribution is determined by
\begin{equation}
\begin{aligned}
2 \abs{\mathcal{Z} } =&
\left\{
\left[J'(1+ \tilde{\vb{n}}_{1} \cdot \tilde{\vb{n}}_{2})- \vb{M}\cdot(\tilde{\vb{n}}_{1} +\tilde{\vb{n}}_{2})\right]^2 \right. \\
 & \quad \left. +\left[\vb{M}\cdot(\tilde{\vb{n}}_{1} \cross \tilde{\vb{n}}_{2})\right]^2 \right\}^{\frac{1}{2}};
\end{aligned}
\end{equation}
The sign of $J'$ enters the competition between the two exchange sectors; a relative phase of $\pi$ does not reverse the oriented spiral normal. 
A circular state in a pure exchange sector reaches the fixed-$\bQ$ LT bound, as follows. Introduce
\begin{equation}
\chi_+=\frac{\widetilde  \psi^{(1)} + \widetilde \psi^{(2)}}{\sqrt2},
\qquad
\chi_-=\frac{\widetilde \psi^{(1)}-\widetilde  \psi^{(2)}}{\sqrt2},
\end{equation}
the energy is decomposed into two sectors: 
\begin{equation}
E_{\text{2-sub}}=\chi_+^\dagger(H_0 + K_g)\chi_+
+\chi_-^\dagger(H_0 - K_g)\chi_-.
\end{equation}
The minimum eigenvalues of the Hermitian sector matrices give the bound 
\begin{equation}
E_{\text{2-sub}} \ge e_+\norm{ \chi_+}^2+e_-\norm{ \chi_-}^2,
\end{equation}
where 
\begin{equation}
	e_{s} = J +s J' - \norm{\vb D + s \vb M} \quad (\text{for } s=\pm1)
\end{equation}
is the lowest eigenvalue of 
\begin{equation}
	H_{s} = H_0 + s K_g = (J + s J' )\mathbb{I} + i [ \vb D + s \vb M]_{\cross}. \label{eq:H_pm}
\end{equation}
Suppose, for example, that \(e_+<e_-\). Taking the normalization condition that $ \norm{ \chi_+}^2+\norm{\chi_-}^2 =\norm{\psi^{(1)}}^2+\norm{\psi^{(2)}}^2=2$, then
$E\ge 2e_+ +(e_--e_+)\|\chi_-\|^2. $
Any nonzero \(\chi_-\) raises this lower bound. If the energy \(2e_+\) is physically feasible, the ground state must consequently have $\chi_-=0$, which implies $\widetilde\psi^{(2)}=\widetilde \psi^{(1)}$. 
Similarly, if $e_-<e_+ $, obtaining the lowest energy requires $\chi_+=0$, implying $\widetilde\psi^{(2)}=-\widetilde \psi^{(1)}$. 
As shown in Sec.~\ref{sec:one_sub_theory}, each $H_s$ has a lowest eigenvector that is a normalized circular polarization. Setting $\widetilde\psi^{(2)}=s \widetilde\psi^{(1)}$, which therefore satisfies both local constraints, gives $E=2e_s$.
The fixed-$\bQ$ minimum for isotropic symmetric interactions in this transformed basis is therefore
\begin{enumerate}[label=(\roman*)]
    \item $ e_+ <e_-:$
    \begin{equation}
    	\widetilde \psi^{(2)}=\widetilde \psi^{(1)}, \quad \tilde{\vb{n}}_{1}=\frac{\vb D + \vb M}{\norm{\vb D + \vb M} }.
    \end{equation}
    \item $ e_- <e_+ :$
    \begin{equation}
    	\widetilde \psi^{(2)}=-\widetilde \psi^{(1)}, \quad \tilde{\vb{n}}_{1}=\frac{\vb D - \vb M}{\norm{\vb D - \vb M} }.
    \end{equation}
\end{enumerate}
If $\vb D+s\vb M=0$ in the selected sector, its normal is arbitrary and the corresponding normalized-vector expression is not used. At $e_+=e_-$, both pure sectors reach the minimum; mixtures can also be minima if they satisfy the local constraints.

\subsubsection{Generic exchange-sector formulas}
\label{sec:two_sub_sectors}
For generic anisotropic exchange in a sublattice-exchange system, the unrestricted two-sublattice problem remains a coupled optimization over two normals. 
Here we provide the solutions for the two sublattice-exchange sectors: $\widetilde\psi^{(2)}=\pm \widetilde \psi^{(1)}$. 

Following the above construction, for arbitrary real symmetric $\mathcal R$ and $\mathcal K$, we define 
\begin{equation}
 R_s=\mathcal R+s\mathcal K,\qquad
 \vb d_s=\bD+s\vb M,\qquad H_s=R_s+i[\vb d_s]_{\cross}.
\end{equation}
Within sector $s=\pm1$, $E_s=\chi^\dagger H_s\chi$ with $\chi^\dagger\chi=2$ and $\chi^\intercal \chi=0$. Thus the full anisotropic branch reduces exactly to the one-sublattice problem discussed in Sec. \ref{sec:one_sub_theory}. A compact, basis-independent secular equation is
\begin{equation}
\vb d_s^\intercal(\lambda_s\mathbb I-R_s)^{-2}\vb d_s=1,
 \label{eq:sector_secular}
\end{equation}
where $\lambda_s>\lambda_{\max}(R_s)$ for a regular minimum. Singular minima at $\lambda_s=\lambda_{\max}(R_s)$ must be treated separately, as in the one-sublattice problem. 
For the regular solution,
\begin{align}
 \tilde{\vb n}_s&=(\lambda_s\mathbb I-R_s)^{-1}\vb d_s,\nonumber\\
 E_s^{\min}&=\mathrm{Tr}R_s-\lambda_s
 -\vb d_s^\intercal(\lambda_s\mathbb I-R_s)^{-1}\vb d_s.
 \label{eq:sector_energy}
\end{align}
We note that Eqs.~(\ref{eq:sector_secular})--(\ref{eq:sector_energy}) provide exact branch minima; however, they do not guarantee a universal global solution of the anisotropic two-sublattice problem.

\subsection{General anisotropy and simultaneous symmetry constraints}
\label{app:anisotropy_details}
For the full model (\ref{eq:preserve_model}), Eq.~(\ref{eq:preserve_branch_energy}) remains exact within the invariant circular branches, but tilted solutions require minimizing Eq.~(\ref{eq:E2sub_final}) on $S^2\times S^2$. By Eq.~(\ref{eq:preserve_blocks}), the phase-minimized energy obeys
\begin{equation}
 E(\vu n_1,\vu n_2)=E(\mathcal D_g\vu n_1,\mathcal D_g\vu n_2).
 \label{eq:preserve_energy_symmetry}
\end{equation}
Both normals transform together; there is generally no invariance under transforming only one. Any symmetry-breaking minimum has an equally energetic symmetry-related partner, although additional degeneracies may enlarge the family.

If two independent operations from Table~\ref{tab:one_little_three} both preserve the sublattices, their constraints must be intersected. In the $(\vu e_\parallel,\vu e_\perp,\vu e_z)$ basis this makes each $\mathcal H_{\mu\mu}$ real diagonal, forces $\vb D^{(\mu)}=0$, and makes $\mathcal H_{12}$ complex diagonal. The energy remains invariant under the corresponding simultaneous transformations of both normals. Nevertheless, no circular spiral can preserve both operations up to phase: their axial $+1$ eigenspaces are distinct one-dimensional axes and have no common unit vector. This is an instance in which the symmetry of the Hamiltonian cannot be imposed on every hard-spin minimum.

\section{Point-group constraints for one magnetic sublattice}
\label{app:pointgroups_one}

\subsection{Arm-preserving symmetries for \texorpdfstring{$\mathbf Q=(q,q,0)$}{Q=(q,q,0)}}
\label{sec:one_sub_diagonal}

By the symmetry constraints in Eq.~(\ref{eq:one_little_constraints}), the forms of Hamiltonian under $M_{1\bar10}$, $M_z$, and $C_{2[110]}$ are summarized in Table~\ref{tab:one_little_three}. 
Each row is an independent constraint; if several operations are present, their constraints must be intersected.
\begin{table*}[t]
\caption{Arm-preserving one-sublattice symmetries for $\bQ=(q,q,0)$. The axial matrices are expressed in the basis $(\hat{\vb e}_\parallel,\hat{\vb e}_\perp,\hat{\vb e}_z)$ of Eq.~(\ref{eq:diagonal_exchange_basis}). Entries of $\mathcal D_g$ refer to the basis in Eq.~(\ref{eq:diagonal_exchange_basis}). Only the listed off-diagonal entries of the real symmetric $\mathcal R$ must vanish.}
\label{tab:one_little_three}
\begin{ruledtabular}
\begin{tabular}{llll}
Operation & $\mathcal D_g$ & Vanishing entries of $\mathcal R$ & Allowed $\bD$\\
\hline
$M_{1\bar10}$ & $\mathrm{diag}(-1,+1,-1)$ & $\mathcal R_{\parallel\perp},\mathcal R_{\perp z}$ & $D_\perp\hat{\vb e}_\perp$\\
$M_z$ & $\mathrm{diag}(-1,-1,+1)$ & $\mathcal R_{\parallel z},\mathcal R_{\perp z}$ & $D_z\hat{\vb e}_z$\\
$C_{2[110]}$ & $\mathrm{diag}(+1,-1,-1)$ & $\mathcal R_{\parallel\perp},\mathcal R_{\parallel z}$ & $D_\parallel\hat{\vb e}_\parallel$\\
\end{tabular}
\end{ruledtabular}
\end{table*}

\subsection{Arm-reversing symmetries for \texorpdfstring{$\mathbf Q=(q,q,0)$}{Q=(q,q,0)}}
\label{sec:one_sub_arm_reversing}

We next consider the arm-reversing point parts $I_{\mathrm{nv}}$, $C_{2[1\bar10]}$, $C_{2z}$, and $M_{110}$, provided they belong to the parent space group. Their axial spin matrices and Hamiltonian constraints are summarized in Table~\ref{tab:one_arm_reversing}, using the same basis as Table~\ref{tab:one_little_three}. The results follow from Eq.~(\ref{eq:one_anti_constraints}): $\mathcal R$ is invariant under conjugation by $\mathcal D_{g'}$, while $\bD$ must lie in its $-1$ eigenspace.

\begin{table*}[t]
\caption{Arm-reversing one-sublattice symmetries for $\bQ=(q,q,0)$. The axial matrices are expressed in the basis $(\hat{\vb e}_\parallel,\hat{\vb e}_\perp,\hat{\vb e}_z)$. Only the listed entries of the real symmetric matrix $\mathcal R$ must vanish; the components of the allowed DM vectors are arbitrary real numbers. }
\label{tab:one_arm_reversing}
\begin{ruledtabular}
\begin{tabular}{llll}
Operation & $\mathcal D_{g'}$ & Vanishing entries of $\mathcal R$ & Allowed $\bD$\\
\hline
$I_{\mathrm{nv}}$ & $\mathrm{diag}(+1,+1,+1)$ & None & $\vb 0$\\
$C_{2[1\bar10]}$ & $\mathrm{diag}(-1,+1,-1)$ & $\mathcal R_{\parallel\perp},\mathcal R_{\perp z}$ & $D_\parallel\hat{\vb e}_\parallel+D_z\hat{\vb e}_z$\\
$C_{2z}$ & $\mathrm{diag}(-1,-1,+1)$ & $\mathcal R_{\parallel z},\mathcal R_{\perp z}$ & $D_\parallel\hat{\vb e}_\parallel+D_\perp\hat{\vb e}_\perp$\\
$M_{110}$ & $\mathrm{diag}(+1,-1,-1)$ & $\mathcal R_{\parallel\perp},\mathcal R_{\parallel z}$ & $D_\perp\hat{\vb e}_\perp+D_z\hat{\vb e}_z$\\
\end{tabular}
\end{ruledtabular}
\end{table*}

For the same axial matrix, arm-preserving and arm-reversing operations impose identical restrictions on $\mathcal R$, but opposite conditions on $\bD$. The former permits the $+1$ eigenspace of $\mathcal D_g$, whereas the latter permits the $-1$ eigenspace. In particular, spin is unchanged by inversion, so $\mathcal D_{I_{\mathrm{nv}}}=\mathbb I$ and inversion forces $\bD=0$ in the one-sublattice model (no DM interaction). The symmetric interaction remains an arbitrary real symmetric matrix; inversion alone does not require isotropic exchange. 

We note that each row in either Table~\ref{tab:one_little_three} or \ref{tab:one_arm_reversing} is an independent constraint, and constraints are intersected when several operations are present. For example, in the system with both $M_{110}$ and $M_{1\bar10}$, the symmetric part $\mathcal R$ at $\bQ$ is diagonal and $\bD = D_\perp\hat{\vb e}_\perp$.

\subsection{Combined tetragonal and cubic constraints}

Tables~\ref{tab:one_little_three} and \ref{tab:one_arm_reversing} give the constraints imposed by individual operations. Here we intersect these constraints for each tetragonal and cubic parent point group at a generic incommensurate $\bQ=(q,q,0)$. We use conventional crystallographic axes and the right-handed basis $(\hat{\vb e}_\parallel,\hat{\vb e}_\perp,\hat{\vb e}_z)$ of Eq.~(\ref{eq:diagonal_exchange_basis}):
\begin{equation}
 \hat{\vb e}_{\parallel}=\frac{(1,1,0)^\intercal}{\sqrt2},\quad
 \hat{\vb e}_{\perp}=\frac{(-1,1,0)^\intercal}{\sqrt2},\quad
 \hat{\vb e}_z=(0,0,1)^\intercal.
\end{equation}
Special commensurate wave vectors, for which additional operations preserve $\bQ$ modulo a reciprocal-lattice vector, are excluded.

For each parent point group $P$, we identify arm-preserving symmetry operations or arm-reversing operations. The possible arm-preserving operations are collected in $P_+$, while arm-reversing ones are in $P_-$, which are 
\begin{equation}
\begin{aligned}
 P_+&=\{E,M_{1\bar10}, M_{z},C_{2[110]}\} , \\
 P_-&= \{I_{\mathrm{nv}},C_{2[1\bar10]},C_{2z},M_{110} \},
\end{aligned}
 \label{eq:app_actual_arm_sets}
\end{equation}
respectively.
The operation sets in the tables below are subsets of the maximal lists introduced in Sec.~\ref{sec:one_sub_little}. To shorten the tables, write
\begin{equation}
\begin{aligned}
 m_\perp&=M_{1\bar10},& m_\parallel&=M_{110},\\
 c_\parallel&=C_{2[110]},&c_\perp&=C_{2[1\bar10]},\\
 c_z&=C_{2z},&\mathcal I&=I_{\mathrm{nv}}.
\end{aligned}
\label{eq:app_op_notation}
\end{equation}
The mirror subscripts specify their normal directions. Every listed parent group contains $c_z$, which reverses $\bQ$. Consequently,
\begin{equation}
 P_-=c_zP_+,\qquad
 c_zm_z=\mathcal I,\quad c_zm_\perp=m_\parallel,
 \quad c_zc_\parallel=c_\perp.
 \label{eq:app_coset}
\end{equation}
Thus it is sufficient to impose all arm-preserving constraints and the constraint from $c_z$. The remaining arm-reversing constraints follow by composition.

Writing $\mathcal H=\mathcal R+i[\bD]_{\cross}$, the conditions to be intersected are
\begin{equation}
\begin{aligned}
 \mathcal D_g\mathcal R\mathcal D_g^T&=\mathcal R
 &&(g\in P_+\cup P_-),\\
 \mathcal D_g\bD&=\bD &&(g\in P_+),\\
 \mathcal D_g\bD&=-\bD &&(g\in P_-).
\end{aligned}
\label{eq:app_intersection}
\end{equation}
For example, $4mm$ contains the arm-preserving mirror $m_\perp$, which sets $\mathcal R_{\parallel\perp}=\mathcal R_{\perp z}=0$ and allows only $D_\perp$. The arm-reversing $c_z$ additionally sets $\mathcal R_{\parallel z}=0$. Their intersection gives a diagonal $\mathcal R$ and $\bD=D_\perp\hat{\vb e}_\perp$. This yields form D below.

\begin{table*}[t]
\caption{One-sublattice Hamiltonian constraints for tetragonal parent point groups at generic $\bQ=(q,q,0)$. The operation notation is defined in Eq.~(\ref{eq:app_op_notation}). Only the listed off-diagonal entries of $\mathcal R$ are allowed; all three diagonal entries remain independent. The symbols $\bar42m$ and $\bar4m2$ denote the two crystallographic orientations of the same abstract class $D_{2d}$.}
\label{tab:app_one_tetragonal}
\begingroup\scriptsize\setlength{\tabcolsep}{3pt}
\begin{ruledtabular}
\begin{tabular}{llllll}
Parent group & $P_+$ & $P_-$ & Allowed off-diagonal $\mathcal R$ & Allowed $\bD$ & Form\\
\hline
$4$ & $\{E\}$ & $\{c_z\}$ & $\mathcal R_{\parallel\perp}$ & $D_\parallel\hat{\vb e}_\parallel+D_\perp\hat{\vb e}_\perp$ & A\\
$\bar4$ & $\{E\}$ & $\{c_z\}$ & $\mathcal R_{\parallel\perp}$ & $D_\parallel\hat{\vb e}_\parallel+D_\perp\hat{\vb e}_\perp$ & A\\
$4/m$ & $\{E,m_z\}$ & $\{c_z,\mathcal I\}$ & $\mathcal R_{\parallel\perp}$ & $\vb0$ & B\\
$422$ & $\{E,c_\parallel\}$ & $\{c_z,c_\perp\}$ & None & $D_\parallel\hat{\vb e}_\parallel$ & C\\
$4mm$ & $\{E,m_\perp\}$ & $\{c_z,m_\parallel\}$ & None & $D_\perp\hat{\vb e}_\perp$ & D\\
$\bar42m$ & $\{E,m_\perp\}$ & $\{c_z,m_\parallel\}$ & None & $D_\perp\hat{\vb e}_\perp$ & D\\
$\bar4m2$ & $\{E,c_\parallel\}$ & $\{c_z,c_\perp\}$ & None & $D_\parallel\hat{\vb e}_\parallel$ & C\\
$4/mmm$ & $\{E,m_z,m_\perp,c_\parallel\}$ & $\{c_z,\mathcal I,m_\parallel,c_\perp\}$ & None & $\vb0$ & E\\
\end{tabular}
\end{ruledtabular}\endgroup
\end{table*}

\begin{table*}[t]
\caption{One-sublattice Hamiltonian constraints for cubic parent point groups at generic $\bQ=(q,q,0)$, with the conventions of Table~\ref{tab:app_one_tetragonal}. Cubic symmetry does not equate the three diagonal entries at this fixed nonzero wave vector.}
\label{tab:app_one_cubic}
\begingroup\scriptsize\setlength{\tabcolsep}{3pt}
\begin{ruledtabular}
\begin{tabular}{llllll}
Parent group & $P_+$ & $P_-$ & Allowed off-diagonal $\mathcal R$ & Allowed $\bD$ & Form\\
\hline
$23$ & $\{E\}$ & $\{c_z\}$ & $\mathcal R_{\parallel\perp}$ & $D_\parallel\hat{\vb e}_\parallel+D_\perp\hat{\vb e}_\perp$ & A\\
$m\bar3$ & $\{E,m_z\}$ & $\{c_z,\mathcal I\}$ & $\mathcal R_{\parallel\perp}$ & $\vb0$ & B\\
$432$ & $\{E,c_\parallel\}$ & $\{c_z,c_\perp\}$ & None & $D_\parallel\hat{\vb e}_\parallel$ & C\\
$\bar43m$ & $\{E,m_\perp\}$ & $\{c_z,m_\parallel\}$ & None & $D_\perp\hat{\vb e}_\perp$ & D\\
$m\bar3m$ & $\{E,m_z,m_\perp,c_\parallel\}$ & $\{c_z,\mathcal I,m_\parallel,c_\perp\}$ & None & $\vb0$ & E\\
\end{tabular}
\end{ruledtabular}\endgroup
\end{table*}

We construct the tables for tetragonal parent point groups in Table~\ref{tab:app_one_tetragonal} and the table for cubic parent point groups in Table~\ref{tab:app_one_cubic}, in each table it shows for each point group the arm-preserving symmetries (in $P_+$) and arm-preserving symmetries (in $P_-$) and the allowed Hamiltonian matrix elements. 
There are five resulting (Hermitian) Hamiltonian forms:
\begin{align}
\mathcal H_{\mathrm A}&=\begin{pmatrix}
r_\parallel&u&iD_\perp\\u&r_\perp&-iD_\parallel\\-iD_\perp&iD_\parallel&r_z
\end{pmatrix},\label{eq:app_HA}\\
\mathcal H_{\mathrm B}&=\begin{pmatrix}
r_\parallel&u&0\\u&r_\perp&0\\0&0&r_z
\end{pmatrix},\label{eq:app_HB}\\
\mathcal H_{\mathrm C}&=\begin{pmatrix}
r_\parallel&0&0\\0&r_\perp&-iD_\parallel\\0&iD_\parallel&r_z
\end{pmatrix},\label{eq:app_HC}\\
\mathcal H_{\mathrm D}&=\begin{pmatrix}
r_\parallel&0&iD_\perp\\0&r_\perp&0\\-iD_\perp&0&r_z
\end{pmatrix},\label{eq:app_HD}\\
\mathcal H_{\mathrm E}&=\mathrm{diag}(r_\parallel,r_\perp,r_z).
\label{eq:app_HE}
\end{align}
Here $r_\alpha=\mathcal R_{\alpha\alpha}$, $u=\mathcal R_{\parallel\perp}$, and all parameters are real functions of $\mathbf q$. The real symmetric coefficients are even under $\mathbf q\mapsto -\mathbf q$, while the DM components are odd. Operations taking $\bQ$ outside $\{\bQ,-\bQ\}$ relate these matrices to those at other arms, rather than adding independent local constraints at this $\bQ$. These tables specify allowed Hamiltonians, not the symmetry retained by an energy-minimizing spiral.

\section{Point-group constraints for two magnetic sublattices}
\label{app:pointgroups_two}

\subsection{Derivation of arm-reversing block constraints}
\label{sec:two_sub_arm_reversing}
We now derive the Hamiltonian constraints imposed by arm-reversing symmetry operations, distinguishing operations that preserve the sublattices from those that exchange them. These constraints supplement the arm-preserving constraints when both types of operation are present in the parent symmetry group. Table~\ref{tab:two_arm_block_constraints} summarizes the four combinations of Fourier-arm and sublattice actions.

An operation that reverses $\bQ$ may either preserve or exchange the magnetic sublattices. These two actions must be distinguished when extending the one-sublattice constraints of Sec.~\ref{sec:one_sub_arm_reversing}. We retain the position-based Fourier convention of Eq.~(\ref{eq:realspace_reconstruction}) and assume a generic incommensurate $\bQ$ with $R_g\bQ=-\bQ$ exactly. The spatial operation acts on the amplitude as
\begin{equation}
\Psi\mapsto\mathcal U_g\Psi^*,\qquad
\mathcal U_g=e^{-i\bQ\cdot\boldsymbol\tau_g}
(P_g\otimes\mathcal D_g),
\label{eq:two_arm_reverse_action}
\end{equation}
where $P_g$ is the two-site permutation matrix and $\mathcal D_g=\det(R_g)R_g$ acts on axial spins. Conjugation originates from the reality relation between the Fourier arms, not from time reversal~\cite{PerezMato2012,YambeHayami2022}. Hamiltonian invariance requires
\begin{equation}
\mathcal H(\bQ)=\mathcal U_g\mathcal H(\bQ)^*\mathcal U_g^\dagger.
\label{eq:two_arm_reverse_H}
\end{equation}
The common translation phase cancels from this condition. Basis-dependent phase factors must instead be retained if a different Fourier gauge is used.

\subsubsection{Preserved and exchanged sublattices}
For $P_g=I_2$, Eq.~(\ref{eq:two_arm_reverse_H}) gives
\begin{equation}
\begin{aligned}
\mathcal H_{\mu\mu}&=\mathcal D_g \mathcal H_{\mu\mu}^*\mathcal D_g^\intercal,\\
\mathcal H_{12 }&=\mathcal D_g \mathcal H_{12 }^*\mathcal D_g^\intercal.
\end{aligned}
\label{eq:two_reverse_preserve_blocks}
\end{equation}
Thus $\mathcal R^{(\mu)}$ is invariant under conjugation by $\mathcal D_g$, while $\mathcal D_g\vb D^{(\mu)}=-\vb D^{(\mu)}$. Each diagonal block obeys the one-sublattice table independently; the additional condition on $\mathcal H_{12 }$ is essential.

For $P_g=\mqty(0&1\\1&0)$, the corresponding constraints are
\begin{equation}
\begin{aligned}
\mathcal H_{11}&=\mathcal D_g \mathcal H_{22}^*\mathcal D_g^\intercal,\\
\mathcal H_{22}&=\mathcal D_g \mathcal H_{11}^*\mathcal D_g^\intercal,\\
\mathcal H_{12 }&=\mathcal D_g \mathcal H_{12 }^\intercal \mathcal D_g^\intercal.
\end{aligned}
\label{eq:two_reverse_exchange_blocks}
\end{equation}
The transpose in the last line follows from $(\mathcal H_{21})^*=(\mathcal H_{12 }^\dagger)^*=\mathcal H_{12 }^\intercal$. In particular,
\begin{equation}
\mathcal R^{(2)}=\mathcal D_g\mathcal R^{(1)}\mathcal D_g^\intercal,
\qquad \vb D^{(2)}=-\mathcal D_g\vb D^{(1)}.
\label{eq:two_reverse_exchange_DM}
\end{equation}
Unlike arm-preserving exchange, arm reversal introduces a minus sign in the relation between the two antisymmetric fields.

For $\bQ=(q,q,0)$, the axial matrices for $I_{\rm nv}$, $C_{2[1\bar10]}$, $C_{2z}$, and $M_{110}$ are listed in Table~\ref{tab:one_arm_reversing}. Write $\mathcal D_g=\mathrm{diag}(\eta_\parallel,\eta_\perp,\eta_z)$ in that table's basis. An arm-reversing operation preserving both sublattices requires ($X=\mathcal H_{12}$)
\begin{equation}
X_{ab}=\eta_a\eta_bX_{ab}^*.
\label{eq:reverse_X_entries_preserve}
\end{equation}
Entries with $\eta_a\eta_b=+1$ are real; those with $\eta_a\eta_b=-1$ are purely imaginary. An exchanging operation instead requires
\begin{equation}
X_{ab}=\eta_a\eta_bX_{ba}.
\label{eq:reverse_X_entries_exchange}
\end{equation}
This is a symmetry or antisymmetry relation without conjugation. In particular, $X_{aa}$ may be complex. Neither condition generally makes $X$ Hermitian.

Inversion has $\mathcal D_{I_\mathrm{nv}}=\mathbb I$. If it preserves each sublattice, the diagonal blocks and $X$ are real, so $\vb D^{(1)}=\vb D^{(2)}=0$. If inversion exchanges the sublattices, however,
\begin{equation}
\mathcal H_{22}=\mathcal H_{11}^*,\qquad \mathcal H_{12}=\mathcal H_{12}^\intercal,\qquad
\vb D^{(2)}=-\vb D^{(1)}.
\label{eq:inversion_exchange_blocks}
\end{equation}
Nonzero opposite intra-sublattice antisymmetric fields are then allowed. If $\mathcal H_{11}=\mathcal H_{22}$ is imposed independently, both fields must vanish. Inversion alone does not require isotropic symmetric exchange.

\subsection{Site permutations and parent-group compatibility}
We assume exactly two magnetic basis sites per primitive cell, forming a closed set under the parent space group. A Bravais translation preserves each basis label. 
As an arm-preserving or arm-reversing operation can be either sublattice-preserving or sublattice-exchanging, even for a group it might allow multiple possible Hamiltonian forms. 
However, the site permutations associated with different symmetry operations are not independent, since they must satisfy the group multiplication rules.
Assigning characters to operations in a group as
\begin{equation}
\begin{aligned}
\epsilon_g&=\begin{cases}+1,&\text{sites preserved},\\-1,&\text{sites exchanged},\end{cases}\\
\epsilon_{gh}&=\epsilon_g\epsilon_h.
\end{aligned}
\label{eq:app_site_character}
\end{equation}
this defines a homomorphism from the parent point group to $S_2$. 
These signs describe site permutations; different signs cannot be assigned independently to all operations. Fractional translations contribute Fourier phases but do not invalidate Eq.~(\ref{eq:app_site_character}); two space-group operations with the same point part differ by a Bravais translation and induce the same site permutation.

In all the tetragonal groups, $C_{2z}\in P_-$ is the square of a fourfold rotation ($C_{2z}=C_{4z}^2$) or rotoinversion. Any permutation of two objects squares to the identity, so $C_{2z}$ can only be sublattice-preserving. The same argument applies to the cubic groups $432$ and $m\bar3m$. In the tetrahedral subgroup $23$, the three coordinate-axis twofold rotations are conjugate, so their permutation signs are equal. Since $C_{2x}C_{2y}=C_{2z}$, this common sign obeys $\epsilon^2=\epsilon$ and must be $+1$. This also establishes the result for $m\bar3$ and $\bar43m$. Hence
\begin{equation}
	\epsilon_{c_z}=+1
\label{eq:app_cz_preserves}
\end{equation}
for every tetragonal or cubic parent group listed here. 

Every reversing operation has the form $c_zh_+$, with $h_+\in P_+$, and therefore has the same site-permutation sign as $h_+$. 
When present, referring to Eq. (\ref{eq:app_coset}),
\begin{equation}
\epsilon_{\mathcal I}=\epsilon_{m_z},\qquad
\epsilon_{m_\parallel}=\epsilon_{m_\perp},\qquad
\epsilon_{c_\perp}=\epsilon_{c_\parallel}.
\label{eq:app_reverse_site_signs}
\end{equation}
For $4/mmm$ and $m\bar3m$, $c_\parallel=m_zm_\perp$ implies
\begin{equation}
\epsilon_{c_\parallel}=\epsilon_{m_z}\epsilon_{m_\perp}.
\label{eq:app_two_signs}
\end{equation}
For example, Table~\ref{tab:app_one_tetragonal} shows that the tetragonal point groups \(4/m\), \(422\), \(4mm\), \(\bar42m\), and \(\bar4m2\) each contain only one nontrivial arm-preserving operation \(h_+\) and two arm-reversing operations, \(c_z\) and \(c_zh_+\). Since \(c_z\) preserves the two sublattices, the other arm-reversing operation has the same site-permutation sign as \(h_+\): $ \epsilon_{c_zh_+}=\epsilon_{c_z}\epsilon_{h_+}=\epsilon_{h_+}$.

Tables~\ref{tab:app_two_tetragonal} and
\ref{tab:app_two_cubic} classify the two-sublattice Hamiltonians
for tetragonal and cubic parent point groups, respectively.
For each group, they identify the independent site-permutation
signs, the number of distinct cases at fixed $\bQ=(q,q,0)$,
and the corresponding Hamiltonian forms.
The arm-reversing site actions follow from
Eq.~(\ref{eq:app_reverse_site_signs}) and therefore require
no additional independent signs. 
The classification assumes exactly two magnetic sites per primitive cell. The actual space group and occupied magnetic Wyckoff positions select the case.

\begin{table*}[t]
\caption{Two-sublattice cases for tetragonal parent point groups. The arm sets are those of Table~\ref{tab:app_one_tetragonal}. Each sign $+$ or $-$ means site preservation or exchange. In every row $\epsilon_{c_z}=+1$; all other reversing signs follow from Eq.~(\ref{eq:app_reverse_site_signs}). The complete block matrices for the listed forms are specified in Table~\ref{tab:app_two_forms} and Eqs.~(\ref{eq:app_X0})--(\ref{eq:app_Xboth}).}
\label{tab:app_two_tetragonal}
\begin{ruledtabular}
\begin{tabular}{llll}
Parent group & Independent local signs & Number of cases & Corresponding forms\\
\hline
$4$ & None & 1 & $\mathcal T_0$\\
$\bar4$ & None & 1 & $\mathcal T_0$\\
$4/m$ & $\epsilon_{m_z}=\pm1$ & 2 & $\mathcal T_{z+},\mathcal T_{z-}$\\
$422$ & $\epsilon_{c_\parallel}=\pm1$ & 2 & $\mathcal T_{a+},\mathcal T_{a-}$\\
$4mm$ & $\epsilon_{m_\perp}=\pm1$ & 2 & $\mathcal T_{b+},\mathcal T_{b-}$\\
$\bar42m$ & $\epsilon_{m_\perp}=\pm1$ & 2 & $\mathcal T_{b+},\mathcal T_{b-}$\\
$\bar4m2$ & $\epsilon_{c_\parallel}=\pm1$ & 2 & $\mathcal T_{a+},\mathcal T_{a-}$\\
$4/mmm$ & $(\epsilon_{m_z},\epsilon_{m_\perp})=(++),(+-),(-+),(--)$ & 4 & $\mathcal T_{++},\mathcal T_{+-},\mathcal T_{-+},\mathcal T_{--}$\\
\end{tabular}
\end{ruledtabular}
\end{table*}

\begin{table*}[t]
\caption{Two-sublattice cases for cubic parent point groups, with the conventions of Table~\ref{tab:app_two_tetragonal}. The arm sets are those of Table~\ref{tab:app_one_cubic}. Full cubic symmetry restricts the site permutations, but does not impose equality of unrelated matrix coefficients at this fixed $\bQ$.}
\label{tab:app_two_cubic}
\begin{ruledtabular}
\begin{tabular}{llll}
Parent group & Independent local signs & Number of cases & Corresponding forms\\
\hline
$23$ & None & 1 & $\mathcal T_0$\\
$m\bar3$ & $\epsilon_{m_z}=\pm1$ & 2 & $\mathcal T_{z+},\mathcal T_{z-}$\\
$432$ & $\epsilon_{c_\parallel}=\pm1$ & 2 & $\mathcal T_{a+},\mathcal T_{a-}$\\
$\bar43m$ & $\epsilon_{m_\perp}=\pm1$ & 2 & $\mathcal T_{b+},\mathcal T_{b-}$\\
$m\bar3m$ & $(\epsilon_{m_z},\epsilon_{m_\perp})=(++),(+-),(-+),(--)$ & 4 & $\mathcal T_{++},\mathcal T_{+-},\mathcal T_{-+},\mathcal T_{--}$\\
\end{tabular}
\end{ruledtabular}
\end{table*}

\begin{table*}[t]
\caption{Complete two-sublattice Hamiltonian forms. Each row means Eq.~(\ref{eq:app_Tblock}) with the displayed blocks. $\mathcal H_{\mathrm L}^{(1)}$ and $\mathcal H_{\mathrm L}^{(2)}$ are independent copies of the one-sublattice matrix form L unless $H_{22}$ is explicitly related to $H_{11}$. The off-diagonal blocks $X$ are given below. $N_{\mathbb R}$ counts independent real parameters at a fixed generic $\mathbf q$.}
\label{tab:app_two_forms}
\begin{ruledtabular}
\begin{tabular}{llllll}
Form & Site signs & $H_{11}$ & $H_{22}$ & $X$ & $N_{\mathbb R}$\\
\hline
$\mathcal T_0$ & No additional generator & $\mathcal H_{\mathrm A}^{(1)}$ & $\mathcal H_{\mathrm A}^{(2)}$ & $X_0$ & 21\\
$\mathcal T_{z+}$ & $\epsilon_{m_z}=+1$ & $\mathcal H_{\mathrm B}^{(1)}$ & $\mathcal H_{\mathrm B}^{(2)}$ & $X_{z+}$ & 13\\
$\mathcal T_{z-}$ & $\epsilon_{m_z}=-1$ & $\mathcal H_{\mathrm A}^{(1)}$ & $ZH_{11}Z$ & $X_{z-}$ & 12\\
$\mathcal T_{b+}$ & $\epsilon_{m_\perp}=+1$ & $\mathcal H_{\mathrm D}^{(1)}$ & $\mathcal H_{\mathrm D}^{(2)}$ & $X_{b+}$ & 13\\
$\mathcal T_{b-}$ & $\epsilon_{m_\perp}=-1$ & $\mathcal H_{\mathrm A}^{(1)}$ & $BH_{11}B$ & $X_{b-}$ & 12\\
$\mathcal T_{a+}$ & $\epsilon_{c_\parallel}=+1$ & $\mathcal H_{\mathrm C}^{(1)}$ & $\mathcal H_{\mathrm C}^{(2)}$ & $X_{a+}$ & 13\\
$\mathcal T_{a-}$ & $\epsilon_{c_\parallel}=-1$ & $\mathcal H_{\mathrm A}^{(1)}$ & $AH_{11}A$ & $X_{a-}$ & 12\\
$\mathcal T_{++}$ & $(\epsilon_{m_z},\epsilon_{m_\perp})=(+,+)$ & $\mathcal H_{\mathrm E}^{(1)}$ & $\mathcal H_{\mathrm E}^{(2)}$ & $X_{++}$ & 9\\
$\mathcal T_{+-}$ & $(\epsilon_{m_z},\epsilon_{m_\perp})=(+,-)$ & $\mathcal H_{\mathrm B}^{(1)}$ & $BH_{11}B$ & $X_{+-}$ & 8\\
$\mathcal T_{-+}$ & $(\epsilon_{m_z},\epsilon_{m_\perp})=(-,+)$ & $\mathcal H_{\mathrm D}^{(1)}$ & $ZH_{11}Z$ & $X_{-+}$ & 8\\
$\mathcal T_{--}$ & $(\epsilon_{m_z},\epsilon_{m_\perp})=(-,-)$ & $\mathcal H_{\mathrm C}^{(1)}$ & $ZH_{11}Z$ & $X_{--}$ & 8\\
\end{tabular}
\end{ruledtabular}
\end{table*}

\subsection{Construction of the block matrices}
By adopting the position-based Fourier convention of Eq.~(\ref{eq:realspace_reconstruction}), the common translation phase cancels from Hamiltonian invariance. Define the axial spin matrices
\begin{equation}
\begin{aligned}
Z&=\mathcal D_{c_z}=\mathcal D_{m_z}
   =\mathrm{diag}(-1,-1,+1), \\
 B&=\mathcal D_{m_\perp}
   =\mathrm{diag}(-1,+1,-1),\\
A&=\mathcal D_{c_\parallel}
   =\mathrm{diag}(+1,-1,-1)=ZB.
\end{aligned}
\label{eq:app_ZBA}
\end{equation}
Here $A,B,Z$ denote spin transformations and are distinct from the Hamiltonian labels $\mathcal H_{\mathrm A},\mathcal H_{\mathrm B},\ldots$.
Every two-sublattice Hamiltonian has the form
\begin{equation}
\mathcal T=\begin{pmatrix}H_{11}&X\\X^\dagger&H_{22}\end{pmatrix}.
\label{eq:app_Tblock}
\end{equation}
By Eq.~(\ref{eq:app_cz_preserves}), all cases satisfy
\begin{equation}
 H_{\mu\mu}=ZH_{\mu\mu}^*Z,\qquad X=ZX^*Z.
 \label{eq:app_universal_Z}
\end{equation}
An arm-preserving operator with axial matrix $F\in\{Z,B,A\}$ adds
\begin{equation}
\begin{array}{ll}
\epsilon_g=+1:&H_{\mu\mu}=FH_{\mu\mu}F,\quad X=FXF,\\[2pt]
\epsilon_g=-1:&H_{22}=FH_{11}F,\quad X=FX^\dagger F.
\end{array}
\label{eq:app_generator_blocks}
\end{equation}
Equations~(\ref{eq:app_universal_Z}) and (\ref{eq:app_generator_blocks}) give the full set of local block constraints; the remaining arm-reversing relations are their products. In particular, $X$ need not be Hermitian.

The most general inter-sublattice block consistent with Eq.~(\ref{eq:app_universal_Z}) is
\begin{equation}
 X_0=\begin{pmatrix}a&b&ic\\d&e&if\\ig&ih&j\end{pmatrix}.
 \label{eq:app_X0}
\end{equation}
The blocks for a single additional arm-preserving generator are
\begin{align}
 X_{z+}&=\begin{pmatrix}a&b&0\\c&d&0\\0&0&e\end{pmatrix},&
 X_{z-}&=\begin{pmatrix}a&b&ic\\b&d&ie\\ic&ie&f\end{pmatrix},
 \label{eq:app_Xz}\\
 X_{b+}&=\begin{pmatrix}a&0&ib\\0&c&0\\id&0&e\end{pmatrix},&
 X_{b-}&=\begin{pmatrix}a&b&ic\\-b&d&ie\\-ic&ie&f\end{pmatrix},
 \label{eq:app_Xb}\\
 X_{a+}&=\begin{pmatrix}a&0&0\\0&b&ic\\0&id&e\end{pmatrix},&
 X_{a-}&=\begin{pmatrix}a&b&ic\\-b&d&ie\\ic&-ie&f\end{pmatrix}.
 \label{eq:app_Xa}
\end{align}
For $4/mmm$ and $m\bar3m$, the four cases give
\begin{equation}
\begin{aligned}
 X_{++}&=\mathrm{diag}(a,b,c),\\
 X_{+-}&=\begin{pmatrix}a&b&0\\-b&c&0\\0&0&d\end{pmatrix},\\
 X_{-+}&=\begin{pmatrix}a&0&ib\\0&c&0\\ib&0&d\end{pmatrix},\\
 X_{--}&=\begin{pmatrix}a&0&0\\0&b&ic\\0&ic&d\end{pmatrix}.
\end{aligned}
\label{eq:app_Xboth}
\end{equation}
All letters in these $X$ matrices are independent real parameters within the indicated matrix. Repeated letters across different cases do not identify their values, nor do they identify parameters in the diagonal blocks. Reality relates opposite wave vectors by $X(-\bQ)=X(\bQ)^*$; thus explicitly real entries are even and explicitly imaginary entries are odd under $\mathbf q\mapsto -\mathbf q$ in this Fourier convention.

As an example, $4mm$ has two local cases. If $m_\perp$ preserves the sites, each diagonal block $H_{\mu\mu}$ has form D and $X=X_{b+}$. If $m_\perp$ exchanges the sites, $H_{11}$ has form A (due to $c_z$), $H_{22}=BH_{11}B$, and $X=X_{b-}$. In the latter case $\mathcal R^{(2)}_{\parallel\perp}=-\mathcal R^{(1)}_{\parallel\perp}$, $D_\parallel^{(2)}=-D_\parallel^{(1)}$, and $D_\perp^{(2)}=D_\perp^{(1)}$. The reversing $m_\parallel$ exchanges the sites in this case, while $c_z$ preserves them. For the centrosymmetric classes, inversion is site preserving when $\epsilon_{m_z}=+1$ and site exchanging when $\epsilon_{m_z}=-1$. Consequently, inversion does not force each DM vector in $H_{\mu\mu}$ to vanish in the exchanging cases.

Table~\ref{tab:app_two_forms} specifies the block structure
of each Hamiltonian form listed in
Tables~\ref{tab:app_two_tetragonal} and
\ref{tab:app_two_cubic}.
It gives the allowed diagonal blocks, the relation between
them when a symmetry exchanges the sublattices, and the
corresponding inter-sublattice block $X$.
The final column counts the independent real parameters at
a fixed generic $q$.
The diagonal-block forms are defined in
Eqs.~(\ref{eq:app_HA})--(\ref{eq:app_HE}), and the explicit
inter-sublattice matrices are given in
Eqs.~(\ref{eq:app_X0})--(\ref{eq:app_Xboth}).

\section{Numerical validation}
\label{app:numerics}

This appendix provides numerical checks of the
constrained single-$Q$ formulation, including the
one-sublattice secular solution, the two-sublattice
wave-vector benchmark, and the anisotropic
exchange-sector comparison. All constrained
minimizations are restricted to circular single-$Q$
states with unit spin length on each sublattice.

\subsection{Anisotropic one-sublattice check}
\label{sec:num_1sub}

We first validate the one-sublattice secular
solution using the parameters of Fig.~\ref{fig:lambda}.
In the principal-axis basis of the symmetric
exchange matrix $\mathcal R$, we choose
\begin{align}
    (h_1,h_2,h_3)&=(-1.6,-0.2,1.5),\nonumber\\
    (D_1,D_2,D_3)&=(0.2,-0.4,0.2).
\end{align}
The largest regular root of
Eq.~(\ref{eq:lambda}) is
$\lambda_*=1.704953$, yielding the optimized
spiral-plane normal
\begin{equation}
    \vu n=
    0.060515\,\vu z_1
    -0.209979\,\vu z_2
    +0.975831\,\vu z_3.
\end{equation}
The normalization residual is
$|\|\vu n\|^2-1|=5.6\times10^{-16}$.

The corresponding constrained energy is
$E_{\rm hard}=-1.148107$, whereas the lowest
eigenvalue of the Hermitian interaction matrix
is $\lambda_{\min}=-1.678072$.
The LT bound is therefore not saturated, with
\begin{equation}
    E_{\rm hard}-\lambda_{\min}=0.529965.
\end{equation}

To quantify the failure of circularity, we
define the residual of a nonzero complex
three-component amplitude $v$ as
\begin{equation}
    r_{\rm circ}(v)
    =\frac{|v^{T}v|}{v^\dagger v}.
    \label{eq:rcirc_def}
\end{equation}
The lowest LT eigenvector has
$r_{\rm circ}=0.932880$, confirming that it
does not satisfy the circularity condition.
An independent multistart minimization over
the unit sphere reproduces the secular energy
within $8.9\times10^{-16}$.
The reconstructed circular amplitude also
satisfies the corresponding constrained
stationarity equations with a relative
residual of $2.5\times10^{-16}$.

\subsection{Two-sublattice wave-vector benchmark}
\label{app:q_benchmark}

We examine the two-sublattice model of
Sec.~\ref{sec:num_2sub}, varying the symmetric
anisotropy scale $\alpha$ and antisymmetric
interaction scale $\beta$ independently over
$[0,1]$. At each parameter pair, the LT ordering
vector is obtained by minimizing the lowest
eigenvalue of the Hermitian interaction matrix.
The constrained ordering vector is determined
by minimizing the phase-reduced circular
single-$Q$ energy over the wave vector and
two independent unit spiral-plane normals.

The isotropic limit provides an analytical
reference. At $\alpha=\beta=0$, the lowest LT
eigenspace admits locally normalized circular
states, and the two minimizations coincide.
For the chosen exchange parameters,
Eq.~(\ref{eq:num_isotropic_Q}) gives
$Q_0/\pi\simeq0.625011$.

At the anisotropic benchmark
$(\alpha,\beta)=(1,1)$ with $c_0=0.45$,
the LT and constrained minima occur at
$Q_{\rm LT}/\pi\simeq0.430886$ and
$Q_{\rm hard}/\pi\simeq0.651176$,
respectively. Their separation is therefore
$(Q_{\rm hard}-Q_{\rm LT})/\pi\simeq0.220291$.
As shown in Fig.~\ref{fig:LT_hardspin_comparison},
the mismatch persists over a finite region
of the parameter space and is predominantly
controlled by symmetric anisotropy for the
chosen interactions.

\begin{figure}[tbp]
    \centering
    \includegraphics[width=0.98\columnwidth]{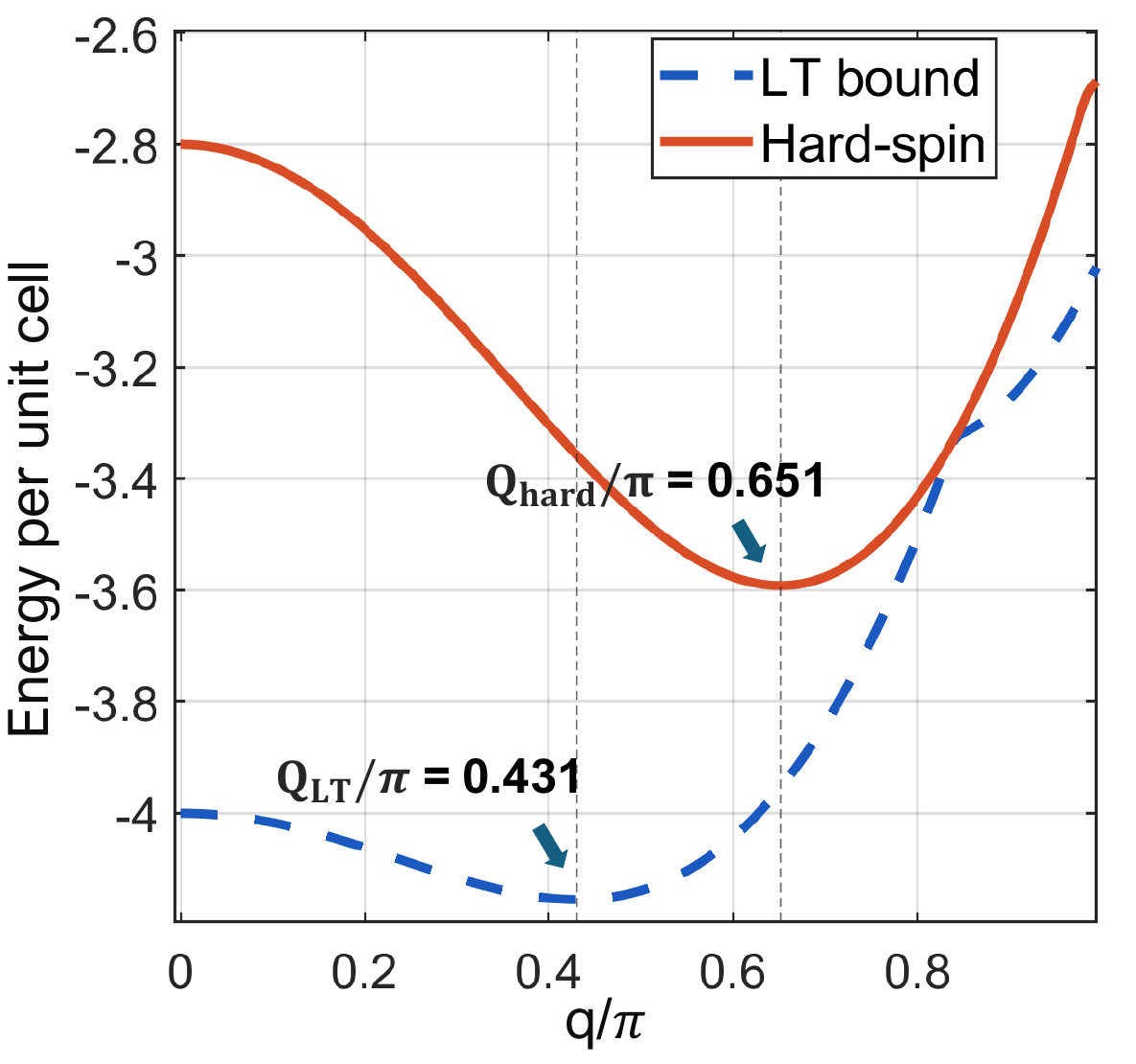}
    \caption{
    LT lower bound $E_{\rm LT}(q)$ and constrained
    circular single-$Q$ energy $E_{\rm hard}(q)$
    for the anisotropic two-sublattice benchmark
    at $(\alpha,\beta)=(1,1)$ with $c_0=0.45$.
    The dashed vertical lines indicate the
    respective minima,
    $Q_{\rm LT}/\pi\simeq0.431$ and
    $Q_{\rm hard}/\pi\simeq0.651$.
    The wave-vector-dependent constraint penalty
    shifts the constrained minimum away from
    the LT prediction. Energies are expressed
    per two-site unit cell in units of $J_1$.
    }
\label{fig:app_LT_dispersion}
\end{figure}

The corresponding constraint penalties
are $\Delta E_{\rm con}(Q_{\rm LT})\simeq0.796509$
and
$\Delta E_{\rm con}(Q_{\rm hard})\simeq0.363176$.
The energy dispersions in
Fig.~\ref{fig:app_LT_dispersion}
demonstrate that this wave-vector-dependent
correction shifts the constrained minimum
away from the LT minimum.

The exchange symmetry of
Eq.~(\ref{eq:num_H}) enforces equal
sublattice weights for nondegenerate
LT eigenvectors, as expressed in
Eq.~(\ref{eq:num_equal_weights}).
Nevertheless, the corresponding circularity
residuals, defined by Eq.~(\ref{eq:rcirc_def}),
are approximately $0.984$ at $Q_{\rm LT}$
and $0.906$ at $Q_{\rm hard}$.
Thus, equal sublattice weights do not
guarantee compatibility with the local
circularity constraints.

\subsection{Two-sublattice numerical consistency}
\label{app:two_sub_validation}

We verify the numerical consistency of the
phase-eliminated formulation for the anisotropic
two-sublattice model of Sec.~\ref{sec:num_2sub},
with $c_0=0.45$ and $(\alpha,\beta)=(1,1)$.
The constrained energy is minimized over
the wave vector and two independent unit
spiral-plane normals using
Eq.~(\ref{eq:E2sub_final}).

The resulting minimizing wave vectors are
$Q_{\rm LT}/\pi\simeq0.430886$ and
$Q_{\rm hard}/\pi\simeq0.651176$.
The corresponding minimum energies per
two-site unit cell are
$E_{\rm LT}\simeq-4.154850$ and
$E_{\rm hard}\simeq-3.591547$.
At the LT minimizing wave vector, the
optimized constrained energy is
$E_{\rm hard}(Q_{\rm LT})\simeq-3.358342$.
At $Q_{\rm hard}$, the optimized circular
state lies in the negative exchange sector.
In the exchange-adapted basis, the two
spiral-plane normals coincide, with one
representative solution given by
\begin{equation}
\widetilde{\vb n}_1
=\widetilde{\vb n}_2
\simeq
-0.340465\,\hat{\vb e}_{\parallel}
+0.940257\,\hat{\vb e}_z.
\end{equation}
The corresponding amplitudes satisfy
$\widetilde\psi^{(2)}
=-\widetilde\psi^{(1)}$.
Consequently, the relative phase is $\pi$
when identical reference frames are chosen
for the two transformed sublattices.

Reconstructing these amplitudes and
evaluating the original quadratic energy gives
\begin{equation}
\begin{aligned}
E_{\rm reduced}
&\simeq-3.591547450369,\\
E_{\rm opt}
&=\widetilde\Psi^\dagger
\widetilde{\mathcal H}(Q_{\rm hard})
\widetilde\Psi\\
&\simeq-3.591547450369.
\end{aligned}
\end{equation}
The agreement verifies the phase elimination
and reconstruction for this stationary state.

The reconstructed complex amplitudes satisfy
the local normalization and circularity
conditions independently on both sublattices,
\begin{equation}
\widetilde\psi^{(\mu)\dagger}
\widetilde\psi^{(\mu)}=1,\qquad
\widetilde\psi^{(\mu)T}
\widetilde\psi^{(\mu)}=0,
\quad \mu=1,2.
\end{equation}
These conditions are stronger than the
global LT normalization constraint.

As an additional consistency check,
the multipliers extracted from
Eqs.~(\ref{eq:eps_extract}) and
(\ref{eq:kappa_extract}) satisfy
\begin{equation}
\begin{aligned}
\varepsilon_1=\varepsilon_2
&\simeq -1.795773725184,\\
|\kappa_1|=|\kappa_2|
&\simeq 0.211207319215.
\end{aligned}
\end{equation}
Their real-energy sum reproduces
$E_{\rm hard}$, while the nonzero anomalous
multipliers reflect the local circularity
constraints. Direct substitution of the
reconstructed stationary amplitudes into
Eq.~(\ref{eq:nonlinear_eigen}) also gives
residuals at the level of floating-point
precision for this representative solution.

These checks establish consistency between
the phase-reduced energy, reconstructed
amplitudes, local constraints, and nonlinear
stationarity equations at the reported
minimum. They do not constitute an analytical
proof of global minimality over the 
circular single-$Q$ manifold.

\subsection{Anisotropic exchange-sector comparison}
\label{app:sector_validation}

We finally describe the calculation underlying
Fig.~\ref{fig:sector_comparison}.
The anisotropic model is defined at fixed
$\bQ$ in the exchange-adapted basis by
\begin{equation}
    \widetilde{\mathcal H}(\eta)
    =
    \begin{pmatrix}
        H_0(\eta)&K_g(\eta)\\
        K_g(\eta)&H_0(\eta)
    \end{pmatrix},
\end{equation}
where
\begin{align}
    H_0(\eta)
    &=J\mathbb I+\eta \mathsf A
      +i[\vb D]_{\cross},\nonumber\\
    K_g(\eta)
    &=J'\mathbb I+\eta \mathsf B
      +i[\vb M]_{\cross}.
\end{align}
Here $\eta$ is real, $\mathsf A$ and $\mathsf B$ are real
symmetric anisotropy tensors, and $\vb D$
and $\vb M$ are real axial vectors.
The matrices $H_0$ and $K_g$ are Hermitian.
The parameters are those specified in
Fig.~\ref{fig:sector_comparison}.

The exchange-sector matrices are
$H_s(\eta)=H_0(\eta)+sK_g(\eta)$,
with $s=\pm1$.
The LT lower bound is calculated as
\begin{equation}
    E_{\rm LT}(\eta)
    =2\min_{s=\pm1}
    \lambda_{\min}[H_s(\eta)].
\end{equation}
For each pure sector, the locally constrained
circular energy is evaluated using the
one-sublattice secular formulation of
Appendix~\ref{sec:two_sub_sectors}.
The lowest pure-sector energy is
\begin{equation}
    E_{\rm sector}^{\rm hard}(\eta)
    =\min_{s=\pm1}E_s^{\rm hard}(\eta).
\end{equation}

The circular hard-spin problem is
evaluated independently by minimizing
over two real unit normals
$\tilde{\vb n}_1,\tilde{\vb n}_2\in S^2$.
With
$\Pi_\mu=\Pi(\tilde{\vb n}_\mu)$,
the phase-minimized energy is
\begin{align}
    \mathcal E_\eta
    ={}&
    \sum_{\mu=1}^{2}
    \operatorname{Tr}
    [\Pi_\mu H_0(\eta)]
    \nonumber\\
    &-2\sqrt{
    \operatorname{Tr}
    [\Pi_1K_g(\eta)\Pi_2K_g(\eta)]
    }.
    \label{eq:app_sector_energy}
\end{align}
Thus,
\begin{equation}
    E_{\rm opt}^{\rm hard}(\eta)
    =\min_{\tilde{\vb n}_1,\tilde{\vb n}_2\in S^2}
    \mathcal E_\eta.
\end{equation}
The projectors enforce circularity and unit
spin length on both sublattices, while the
relative in-plane phase has already been
eliminated analytically.

The three energies obey the variational bounds
\begin{equation}
    E_{\rm LT}(\eta)
    \leq E_{\rm opt}^{\rm hard}(\eta)
    \leq E_{\rm sector}^{\rm hard}(\eta).
    \label{eq:app_sector_bounds}
\end{equation}
The first inequality follows from the LT
relaxation of the local constraints.
The second follows because the pure-sector
states form a subset of the optimal circular
two-sublattice manifold.

For the parameters of
Fig.~\ref{fig:sector_comparison}, the numerical
two-normal solution coincides in energy with
the lowest pure-sector branch at small
anisotropy. Beyond approximately
$\eta_c\simeq0.35$, a lower-energy solution
with nonaligned transformed normals is found.
The corresponding relative angle is
\begin{equation}
    \widetilde\theta_{12}
    =\cos^{-1}
    (\tilde{\vb n}_1\cdot\tilde{\vb n}_2).
\end{equation}
Its apparent discontinuity accompanies the
departure from the pure-sector solution. This behavior is specific to the chosen
interaction parameters. While the LT bound
and the constrained pure-sector minima
are obtained from analytical formulations,
the general two-normal minimum is determined
numerically. The observed departure does not
by itself establish a universal transition
or provide an analytical proof of global
minimality.

An additional feature of Fig.~\ref{fig:sector_comparison} is the contact between the LT bound and the optimal circular hard-spin energy at an isolated anisotropy value $\eta_*$. For the tensors in Eq.~(\ref{eq:anisotropy_tensors}), this occurs when the lowest eigenvalues of the two exchange sectors become degenerate,
\begin{equation}
J+J'-2\eta_*=J-J'-\sqrt{\eta_*^2+(D_\parallel-M_\parallel)^2}.
\label{eq:sector_crossing}
\end{equation}
Defining $\delta=D_\parallel-M_\parallel$, the positive crossing occurs at
\begin{equation}
\eta_*=\frac{4J'+\sqrt{4J'^2+3\delta^2}}{3}.
\end{equation}
For the parameters of Fig.~\ref{fig:sector_comparison}, this gives $\eta_*\simeq0.727$.

The saturation of the LT bound at this point can be understood by constructing a circular state from the degenerate eigenvectors. Let $t=\sqrt{\eta_*^2+\delta^2}$. The normalized lowest eigenvectors of the two sectors at the crossing can be chosen as
\begin{equation}
\begin{aligned}
\vb w_+&=\hat{\vb e}_\parallel,\\
\vb w_-&=
\sqrt{\frac{t+\eta_*}{2t}}\,\hat{\vb e}_\perp
-i\,\mathrm{sgn}(\delta)
\sqrt{\frac{t-\eta_*}{2t}}\,\hat{\vb e}_z.
\end{aligned}
\label{eq:crossing_eigenvectors}
\end{equation}
Although neither eigenvector is individually circular, their degeneracy allows a superposition without increasing the LT energy. In the exchange-adapted basis, consider

\begin{equation}
\begin{aligned}
\widetilde\psi^{(1)}&=\rho_+\vb w_++i\rho_-\vb w_-,\\
\widetilde\psi^{(2)}&=\rho_+\vb w_+-i\rho_-\vb w_-,
\end{aligned}
\label{eq:crossing_amplitudes}
\end{equation}
where
\begin{equation}
\rho_+^2=\frac{\eta_*}{t+\eta_*},
\qquad
\rho_-^2=\frac{t}{t+\eta_*}.
\end{equation}

Since $\vb w_+^\dagger\vb w_-=0$ and $\vb w_-^{\intercal}\vb w_-=\eta_*/t$, both sublattice amplitudes satisfy

\begin{equation}
\begin{aligned}
\widetilde\psi^{(\mu)\dagger}\widetilde\psi^{(\mu)}
&=\rho_+^2+\rho_-^2=1,\\
\widetilde\psi^{(\mu)\intercal}\widetilde\psi^{(\mu)}
&=\rho_+^2-\frac{\eta_*}{t}\rho_-^2=0.
\end{aligned}
\label{eq:crossing_constraints}
\end{equation}

Thus the constructed state satisfies the local normalization and circularity constraints while lying entirely within the degenerate lowest LT eigenspace. Consequently, at the crossing,
\begin{equation}
E_{\rm LT}(\eta_*)=
E_{\rm opt}^{\rm hard}(\eta_*)
<E_{\rm sector}^{\rm hard}(\eta_*).
\label{eq:crossing_saturation}
\end{equation}
The LT bound is therefore obtained by a sector-mixed circular state, even though neither pure sector can achieve it individually. This isolated saturation illustrates how degeneracy can restore local hard-spin compatibility without requiring the minimizing state to belong to a single exchange sector.

\clearpage
\bibliographystyle{apsrev4-2}
\bibliography{spinmodel}
\end{document}